\documentclass[preprint,aps,prb]{revtex4-1}

\usepackage{graphicx}
\usepackage{bm}
\usepackage{amsmath}
\usepackage{latexsym}
\usepackage{dcolumn}

\newcommand{\rfig}[1]{Fig.~\ref{#1}}

\makeindex

\begin{document}

\title{Anomalous Hall effect and magnetic orderings in nanothick V$_5$S$_8$}

\author{Jingjing Niu}
\altaffiliation{These authors equally contributed to the work.}
\affiliation{State Key Laboratory for Artificial Microstructure and Mesoscopic Physics, Peking University, Beijing 100871, China}
\affiliation{Collaborative Innovation Center of Quantum Matter, Beijing 100871, China}
\author{Qingqing Ji}
\altaffiliation{These authors equally contributed to the work.}
\affiliation{Center for Nanochemistry, Beijing National Laboratory for Molecular Sciences, College of Chemistry and Molecular Engineering, Peking University, Beijing 100871, China}
\author{Baoming Yan}
\affiliation{State Key Laboratory for Artificial Microstructure and Mesoscopic Physics, Peking University, Beijing 100871, China}
\affiliation{Collaborative Innovation Center of Quantum Matter, Beijing 100871, China}
\author{Mingqiang Li}
\affiliation{Electron Microscopy Laboratory, School of Physics, Peking University, Beijing 100871, China}
\author{Peng Gao}
\affiliation{Collaborative Innovation Center of Quantum Matter, Beijing 100871, China}
\affiliation{Electron Microscopy Laboratory, School of Physics, Peking University, Beijing 100871, China}
\author{Zhongfan Liu}
\affiliation{Center for Nanochemistry, Beijing National Laboratory for Molecular Sciences, College of Chemistry and Molecular Engineering, Peking University, Beijing 100871, China}
\author{Yanfeng Zhang}
\email{yanfengzhang@pku.edu.cn}
\affiliation{Center for Nanochemistry, Beijing National Laboratory for Molecular Sciences, College of Chemistry and Molecular Engineering, Peking University, Beijing 100871, China}
\affiliation{Department of Matrials Science and Engineering, College of Engineering, Peking University, Beijing 100871, China}
\author{Dapeng Yu}
\author{Xiaosong Wu}
\email{xswu@pku.edu.cn}
\affiliation{State Key Laboratory for Artificial Microstructure and Mesoscopic Physics, Peking University, Beijing 100871, China}
\affiliation{Collaborative Innovation Center of Quantum Matter, Beijing 100871, China}
\affiliation{Department of Physics, Southern University of Science and Technology of China, Shenzhen 518055, China}

\begin{abstract}
The rise of graphene marks the advent of two-dimensional atomic crystals, which have exhibited a cornucopia of intriguing properties, such as the integer and fractional quantum Hall effects, valley Hall effect, charge density waves and superconductivity, to name a few. Yet, magnetism, a property of extreme importance in both science and technology, remains elusive. There is a paramount need for magnetic two-dimensional crystals. With the availability of many magnetic materials consisting of van der Waals coupled two-dimensional layers, it thus boils down to the question of how the magnetic order will evolve with reducing thickness. Here we investigate the effect of thickness on the magnetic ordering in nanothick V$_5$S$_8$. We uncover an anomalous Hall effect, by which the magnetic ordering in V$_5$S$_8$ down to 3.2 nm is probed. With decreasing thickness, a breakdown of antiferromagnetism is evident, followed by a spin-glass-like state. For thinnest samples, a weak ferromagnetic ordering emerges. The results not only show an interesting effect of reducing thickness on the magnetic ordering in a potential candidate for magnetic two-dimensional crystals, but demonstrate the anomalous Hall effect as a useful characterization tool for magnetic orderings in two-dimensional systems.
\end{abstract}



\maketitle

\section{Introduction}

Starting from graphene, the research in two-dimensional (2D) crystals has exploded into a set of subareas covering a vast class of materials, including hexagonal boron nitride, transition metal dichalcogenides, silicene and phosphorene, etc. \cite{Geim2013,Butler2013,Li2014b}. With a broad spectrum of properties displayed by 2D crystals on hand, it is tempting to construct van der Waals heterostructures by multistacking so that the material functionality can be greatly expanded\cite{Geim2013}. However, magnetism, a property that has been playing an indispensable role in both science and technology, remains elusive in 2D crystals. On one hand, since most 2D crystals are nonmagnetic, studies have mainly been focused on induced magnetism by defects and adatoms\cite{Zhang2007,Wang2009,Tongay2012,Nair2012,Nair2013,Han2014,Gonzalez-Herrero2016}. On the other hand, it has been actively pursued to find 2D crystals that are intrinsically magnetic\cite{Du2016,Park2016,Lee2016,Wang2016}. The natural approach is to start with known three-dimensional (3D) magnetic materials that consist of weakly bonded 2D layers and investigate how the thickness affects the magnetic ordering. In addition, revealing the evolution can deepen our understanding in phase transitions on low-dimensional systems.

However, the experiment is challenging when one tries to characterize the magnetic properties of 2D crystals. This is because common tools for magnetization characterization, such as magnetometer, neutron scattering, magnetic resonance technique, often require a 3D bulk sample. To study the magnetic property of a single 2D crystal, new and simple methods are highly desired. Raman spectroscopy, combined with theoretical calculations, was employed to infer the antiferromagnetic ordering in atomic layers of FePS$_3$\cite{Lee2016,Wang2016}. Recent studies have utilized the Kerr effect to identify ferromagnetic ordering in 2D materials in the monolayer limit\cite{Gong2017,Huang2017}. 

In this work, we study the thickness dependence of the magnetic ordering in V$_5$S$_8$ nanoflakes. An unusual Hall behavior is found. It is convincingly shown that it stems from the anomalous Hall effect (AHE). The effect enables characterization of the magnetic ordering of a single nano-flake. The thickness dependence of the AHE, combined with magnetoresistance (MR), reveals a breakdown of antiferromagnetism (AFM) with reducing thickness. For thinnest flakes, a weak FM emerges. In the transition region, the competition between AFM and FM interactions gives rise to a spin-glass-like state. Our results not only reveal an interesting effect of reducing thickness on the magnetic ordering in a potential candidate of a magnetic 2D crystal, but also demonstrate the anomalous Hall effect as a useful tool for obtaining information on the magnetic ordering in 2D crystals.

\section{Experimental techniques}

V$_5$S$_8$ flakes were grown at ~600$^\circ$C by a chemical vapor deposition method using solid VCl$_3$ and sulfur precursors under a mixed Ar/H$_2$ gas flow. By lowering the evaporation temperature of VCl$_3$ down to 275-300 $^\circ$C and optimizing the location of SiO$_2$/Si substrates at ~6 cm downstream from the VCl$_3$ precursor, V$_5$S$_8$ nanoflakes with thickness less than 10 nm and domain size up to tens of micrometers could be readily synthesized. The V$_5$S$_8$ samples could be further thinned down to 3.2 nm (2.5 unit cell) by peeling off the as-grown nanoflakes onto fresh SiO$_2$/Si substrates. Consequently, two types of thin samples were measured, as-grown and thinned by mechanical exfoliation of thicker flakes. No significant difference was found (see Fig.~S2 and S3 in the Supplemental Material). Samples are very stable in ambient conditions, as the optical contrast and the resistance stayed the same after days of exposure even in thinnest samples. Bulk samples of V$_5$S$_8$ were purchased from The 2D AGE Company. High resolution transmission electron microscopy imaging and electron diffraction experiments was performed in an FEI Tecnai F30 TEM at 300 kV. The atomic force microscopy image was captured on a Bruker Dimension Icon AFM. Standard electron beam lithography was employed to pattern the Hall bar structure and 5 nm Pd/80 nm Au was used for metallization. As samples are metallic and stable, no special care is needed to ensure a good electrical contact($<$ 100 $\Omega$). Four-probe electrical measurements were carried out using a lock-in method in an Oxford variable temperature cryostat. All magnetotransport measurements were performed with the magnetic field perpendicular to the film plane. Magnetization measurements were performed in a Magnetic Property Measurement System by Quantum Design.

\section{Results and discussions}

\subsection{Crystal and magnetic structures of V$_5$S$_8$}

\begin{figure}[htbp]
\includegraphics[width=1\textwidth]{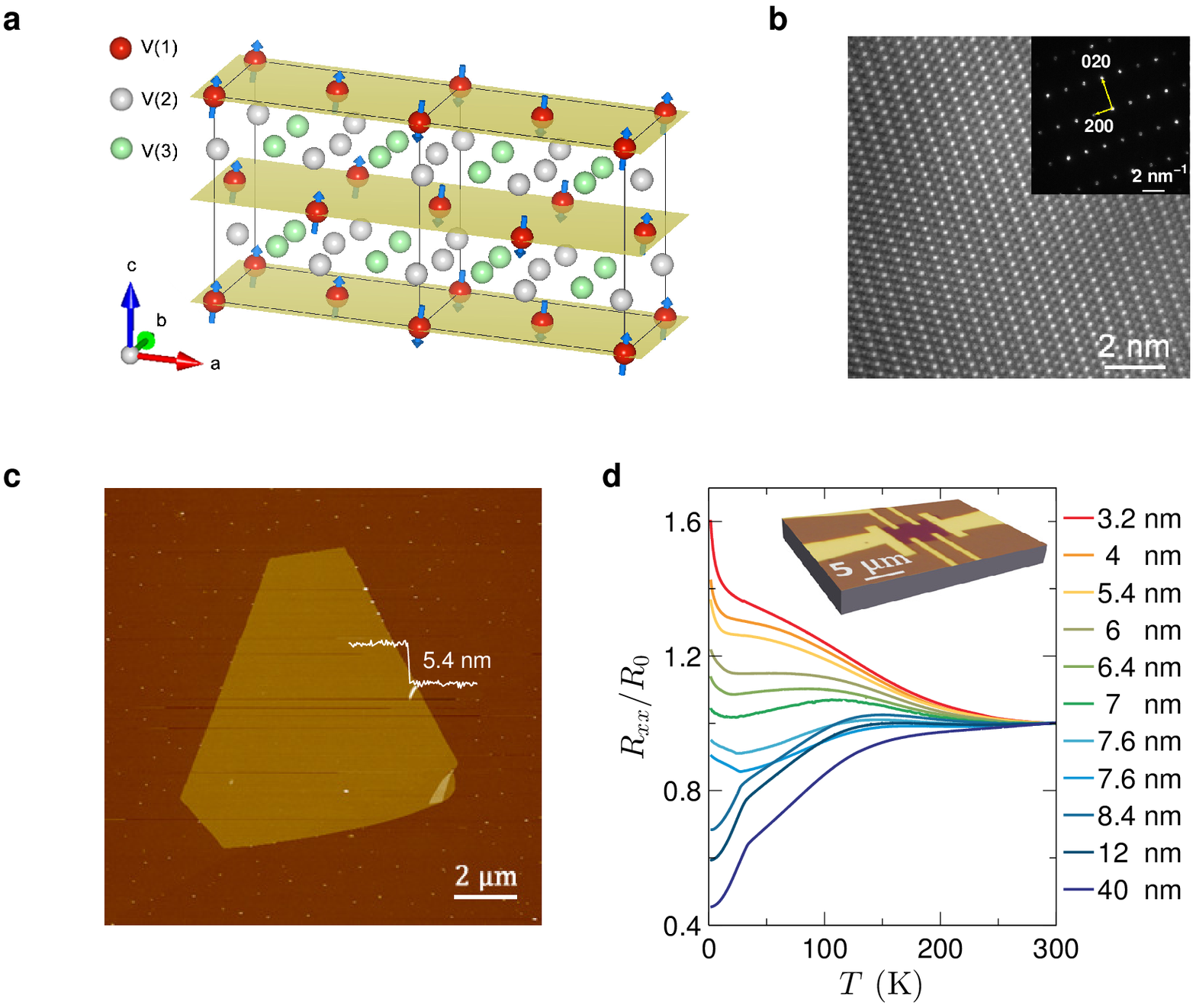}
\caption{Structure, morphology and resistivity of V$_5$S$_8$ flakes. (a) Unit cell of the magnetic structure of V$_5$S$_8$, whose volume is doubled compared with the crystalline unit cell( $a$ is doubled).  (b) High resolution TEM image of a nanosheet with viewing direction of [001]. Inset is the corresponding selected area electron diffraction pattern. (c) Atomic force microscopy image of a 5.4 nm thick flake. The white line shows the height profile across an edge of the flake. (d) Temperature dependence of the resistivity normalized to the resistivity at 300 K, $R/R_{300 K}$, for samples of different thickness. For clarity, only a part of the samples are shown. More data can be found in the Supplemental Material. Inset: A typical Hall bar structure used in the measurements.}
\label{fig1}
\end{figure}

V$_5$S$_8$ is a layered material and can be viewed as VS$_2$ intercalated with V atoms. It has a monoclinic structure ($a=11.396$ $\text\AA$, $b=6.645$ $\text\AA$, $c=11.293$ $\text\AA$, $\alpha=\gamma=90^\circ$, $\beta=91.45^\circ$), space group F2/m, shown in \rfig{fig1}a\cite{Kawada1975}. The VS$_2$ layer is in a distorted $1T$ structure due to V intercalation. The intercalated V atoms are below V atoms in the VS$_2$ layer. Therefore, there are three inequivalent V sites, V(1), V(2) and V(3). Intercalated V atoms are on the V(1) sites, forming a slightly distorted triangular lattice. The magnetic properties of bulk V$_5$S$_8$ are more or less understood\cite{Vries1973,Silbernagel1975,Nozaki1977,Nozaki1978,Kitaoka1980,Funahashi1981,Nakanishi2000}. It is an antiferromagnetic metal below about 32 K. Neutron scattering and nuclear magnetic resonance experiments have suggested that the intercalated V atoms are responsible for the magnetism. Their spins align at 10.4$^\circ$ away from the $c$ axis toward the $a$ axis. The antiferromagnetic alignment of spins is depicted in \rfig{fig1}. The resistivity is metallic and displays a kink at 32 K, which has been identified as the N\'{e}el temperature $T_\text{N}$ (see the Supplemental Material). When a magnetic field is applied parallel to the $c$ axis, a spin-flop (SF) transition occurs at a critical field $H_\text{c}\approx$3.5 T\cite{Nozaki1977,Nakanishi2000}. The AFM ordering persists down to 10 nm\cite{Hardy2016}.

We have studied thin flakes of V$_5$S$_8$ samples with a series of thickness down to 3.2 nm. Monolayer V$_5$S$_8$, consisting of two layers of VS$_2$ and one layer of intercalated V, is 0.847 nm thick and each subsequent layer adds additional 0.565 nm to the thickness. So, 3.2 nm roughly corresponds to five layers. It is worth noting that the interface between the substrate and the 2D material may additionally contribute to the thickness, too. The structure is confirmed by the transmission electron microscopy (TEM) characterization in \rfig{fig1} and Supplemental Material Fig.~S1. A high-resolution image in \rfig{fig1}b shows the structure of a nanoflake V$_5$S$_8$ and the selected area electron diffraction pattern exhibits a rectangular arrangement with $d_{200}=5.75\pm0.05$ $\text\AA$ and $d_{020}=3.35\pm0.05$ $\text\AA$, from which the lattice constants are calculated to be about $a=1.15$ nm and $b=0.67$ nm, in agreement with V$_5$S$_8$\cite{Silbernagel1975,Poddar2002}. However, sometimes we did observe rotation of the $a$ and $b$ axes and the hexagonal lattice of VS$_2$, suggesting the existence of intercalation fluctuations. For thicker flakes, the resistance decreases with temperature and is followed by a sudden drop at 32 K, signaling an AFM transition. The dependence is similar to bulk materials (see the Supplemental Material). Interestingly, when the thickness is below 8.4 nm or so, the resistance drop at 32 K becomes an abrupt increase, maintaining a well defined transition temperature. With further reduction of the thickness, the sharp transition turns into a crossover. Although the low-temperature enhancement of the resistivity is stronger in thinner flakes, it remains relatively low down to 3.2 nm, suggesting an absence of strong localization or opening up of a gap.

\subsection{Magnetotransport and anomalous Hall effect in thick flakes}

\begin{figure}[htbp]
\includegraphics[width=1\textwidth]{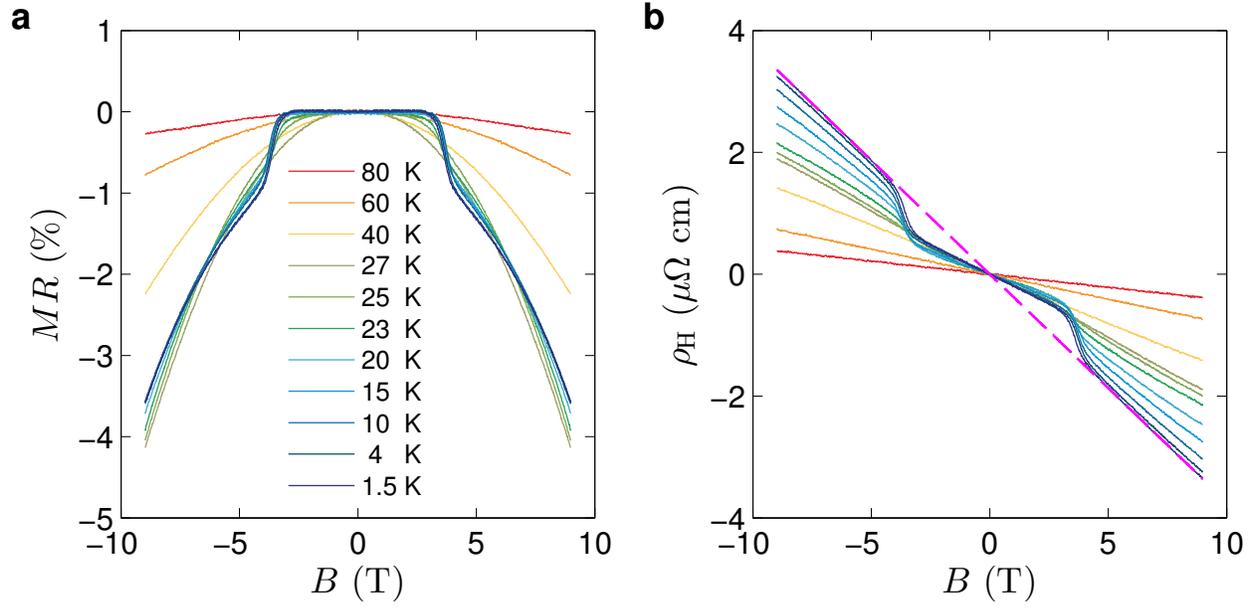}
\caption{Magnetotransport of an 8.4-nm-thick flake. (a) MR at different temperatures. (b) $\rho_\text{H}$ at different temperatures. The magenta dashed line is a linear fit of the high-field data.}
\label{fig2}
\end{figure}

In thicker flakes, the SF transition in the AFM state is clearly manifested in magnetotransport. Figure~\ref{fig2} shows the MR and Hall resistivity $\rho_\text{H}$ for a typical sample (data for more samples can be found in the Supplemental Material). Below $T_\text{N}$, the low field resistance is essentially constant, followed by a sudden decrease at 3.5 T, which results from the SF transition. No appreciable hysteresis has been observed, probably due to a flat barrier between two states on two sides of the transition\cite{Oh2014}. The high field dependence is quadratic and diminishes with temperature, consistent with suppression of scattering from local spin fluctuations in a paramagnetic (PM) state\cite{Yamada1972,Jin2015}. The SF transition indicates a relatively weak spin anisotropy\cite{Groot1986,Balamurugan2014}. In fact, the anisotropy in bulk V$_5$S$_8$ was found to be extremely small\cite{Nozaki1978}.

Accordingly, $\rho_\text{H}$ exhibits an intriguing change of the slope across the SF transition, seen in \rfig{fig2}b. A nonlinear Hall resistivity usually indicates a two-band conduction, but the fact that the high field Hall extrapolates exactly to the origin rules out this possibility, as a simple two-band model cannot reproduce such a behavior. Another explanation would be a field induced change of the carrier density, for instance, breaking down of a spin-density wave state. But, a spin-density wave gaps out a part of the Fermi surface. Thus, its breaking down would recover the gapped Fermi surface, hence increasing the carrier density. The resultant reduction of the Hall resistance is apparently at odds with the experiment, not to mention that no spin-density wave has been reported in the material before.

In magnetic materials, the Hall resistivity consists of two contributions, $\rho_\text{H}=R_0 B+R_\text{AHE}\mu_0 M$, where $R_0$ and $R_\text{AHE}$ are the ordinary and anomalous Hall coefficients, $M$ is the magnetization, and $\mu_0$ the vacuum permeability. Although the AHE often appears in an FM metal, it can also occur in a PM or AFM one\cite{Fert1972,Nakatsuji2015,Luo2015,Suzuki2016}. The difference is that the AHE is linear in $B$ for the latter, as $M$ is also linear in $B$. In our samples, a magnetic field induces a SF transition, which results in an increase of the magnetic susceptibility, hence $\rho_\text{H}$.

To verify this hypothesis, we need to measure $M$, which can be directly measured only for bulk materials. So, the magnetization and transport measurements were carried out for a bulk V$_5$S$_8$ (see Fig.~S4 in the Supplemental Material). A linear relation between $\mu_0 M/B$ and $\rho_\text{H}/B$ was indeed found, confirming the contribution of the AHE (see Fig.~S5 and the discussion in the Supplemental material). Moreover, $R_\text{AHE}$ has a sign opposite to $R_0$. $R_0=0.20$ $\mu\Omega$ cm T$^{-1}$, corresponding to a hole density of 3.13$\times10^{21}$ cm$^{-3}$. The carrier being hole is corroborated by the positive slope of the gate dependence of resistivity for a very thin sample, seen in Supplemental Material Fig.~S9. Apparently, $\rho_\text{H}$ is dominated by negative $R_\text{AHE}$. Therefore, the linear dependence between $R_\text{AHE}$ and $\mu_0 M/B$ provides us with a desperately needed means to gain the information on the magnetization of individual 2D crystals, which is inaccessible, due to their negligible volume, to magnetometers and magnetic resonance techniques. $R_\text{AHE}$ for the bulk is $-162.41$ $\mu\Omega$ cm T$^{-1}$, calculated from the slope in Fig. S5. The longitudinal resistivity $\rho_{xx}$ is around 600 $\mu\Omega$ cm. In this regime, we evaluate the magnitude of the AHE by calculating $S=\mu_0 R_\text{AHE}/\rho_{xx}^2=-0.057$ V$^{-1}$. It is of the same order as in various magnets\cite{Nagaosa2010,Nakatsuji2015}. Being dominated by the AHE, in what follows, the Hall can be simply viewed as the magnetization.

\begin{figure}[htbp]
\includegraphics[width=1\textwidth]{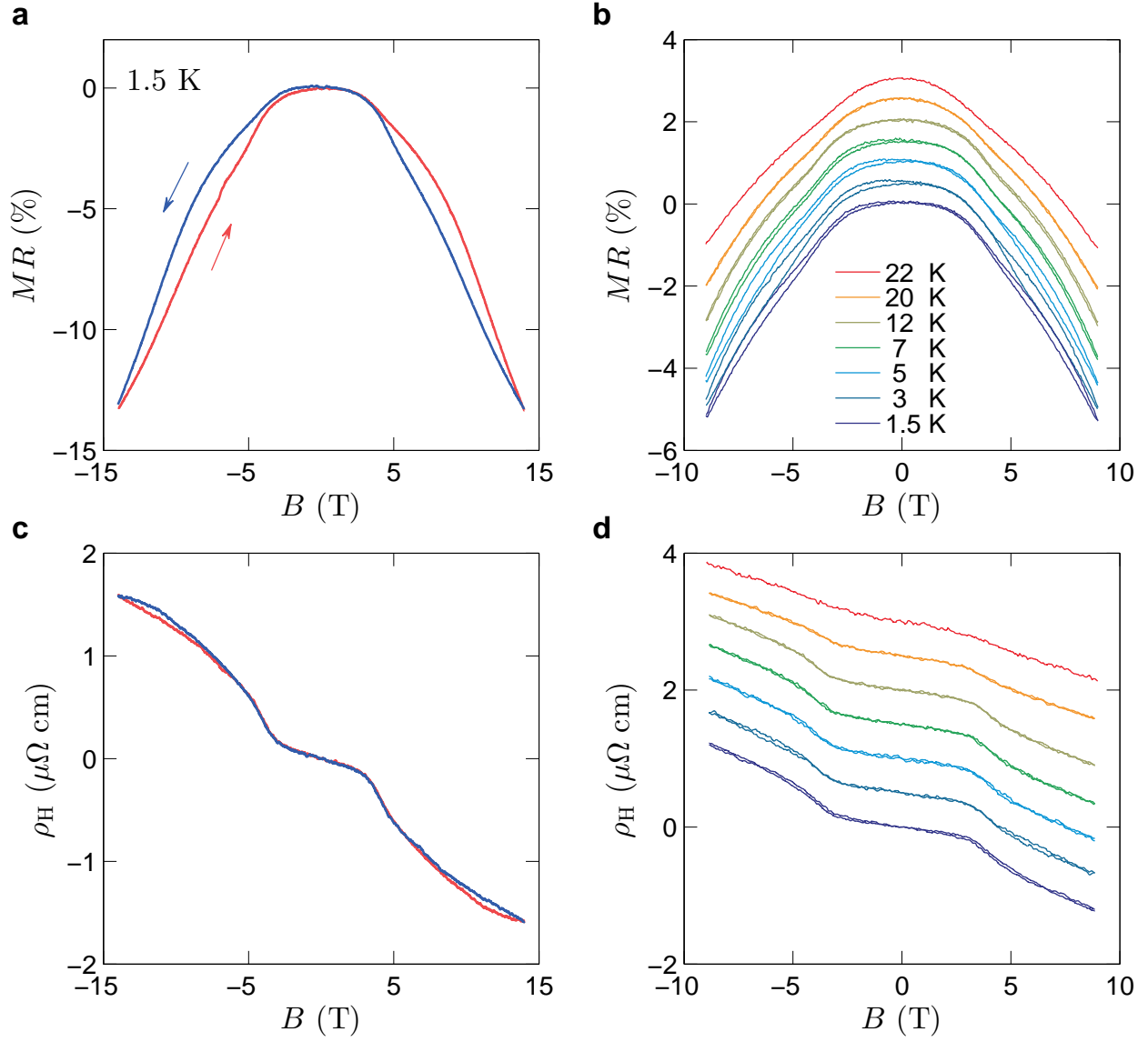}
\caption{Magnetotransport of a 7.6 nm thick flake. (a) MR hysteresis at 1.5 K. (b) MR at different temperatures. (c) $\rho_\text{H}$ hysteresis at 1.5 K. (d) $\rho_\text{H}$ at different temperatures.}
\label{fig3}
\end{figure}

As the thickness is reduced below 8.4 nm, the low temperature resistivity goes up. The SF transition is not as sharp as in thicker flakes. In addition, the magnetotransport becomes hysteretic, suggesting a spin-glass-like state. Typical data for a 7.6 nm thick flake are shown in \rfig{fig3}. The hysteresis is only significant above $H_\text{c}$. Its magnitude increases with the sweeping field and decreases with temperature and eventually disappears at about 12 K. The nonlinear $\rho_\text{H}$, on the other hand, persists to a higher temperature. The hysteresis in MR is more pronounced than that in $\rho_\text{H}$. Since the in-plane magnetization does not contribute to the AHE, but to MR via spin fluctuation scattering, it is speculated that the hysteresis is mainly related to the in-plane spin component.

\subsection{Anomalous Hall effect and ferromagnetic ordering in thin flakes}

\begin{figure}[htbp]
\includegraphics[width=1\textwidth]{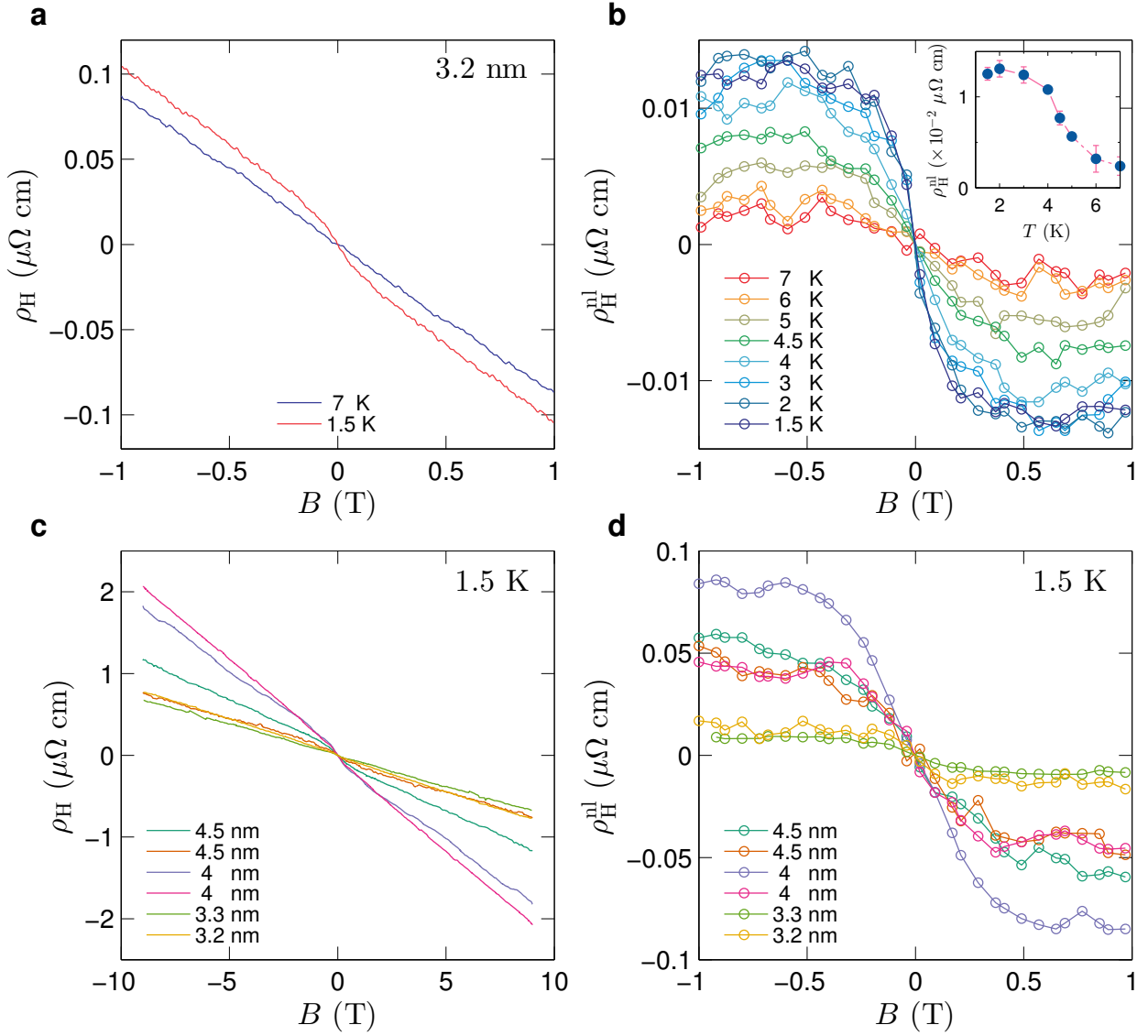}
\caption{$\rho_\text{H}$ of a 3.2 nm thick flake. (a) $\rho_\text{H}$ at 1.5 (red) and 7 (blue) K. (b) Nonlinear Hall resistivity $\rho_\text{H}^\text{nl}$ obtained by subtracting the high field linear background at different temperatures. Inset: Saturation value (averaged over the higher field region) of $\rho_\text{H}^\text{nl}$ as a function of temperature. (c) $\rho_\text{H}$ of other measured thin flakes at 1.5 K. (d) Nonlinear Hall resistivity $\rho_\text{H}^\text{nl}$.}
\label{fig4}
\end{figure}

With further reduction of thickness, the nonlinearity of $\rho_\text{H}$ diminishes. $H_\text{c}$ becomes smaller and the low field slope approaches the high field one. However, when the thickness is below about 5.4 nm, the Hall behavior qualitatively changes. A steep slope emerges in low fields, while it remains linear in high fields, shown in \rfig{fig4}. The high field linear dependence intercepts the $y$ axis at a finite value, in sharp contrast with the zero intercept in thick flakes. After subtracting a high field linear background, the nonlinear part $\rho_\text{H}^\text{nl}$ is extracted and plotted in \rfig{fig4}b. The curves exhibit characteristics of the AHE of an FM metal, suggesting an FM ordering. The saturation value of $\rho_\text{H}^\text{nl}$, which is proportional to the saturation magnetization, decreases with temperature. Its temperature dependence is plotted in the inset of \rfig{fig4}b, from which the Curie temperature can be estimated to be 7 K.

A similar FM type of the AHE has been observed in other thin flakes as well, seen in \rfig{fig4}c and d. No hysteresis has been observed, indicating a rather soft FM, consonant with the weak spin anisotropy indicated by the SF transition as we discussed above. The amplitude of the AHE ranges from 0.01 to 0.1 $\mu\Omega$cm. The fact that the linear AHE is large and non-saturating implies that only a fraction of the magnetic moments participate in the FM ordering, while the rest remain in a PM state. This may be due to residual competition between FM and AFM interactions and inhomogeneity of intercalation. Alternatively, it could be related to the enhanced thermal fluctuation effect in reduced dimensions, which suppresses a long-range order in a Heisenberg spin system, as observed in Cr$_2$Ge$_2$Te$_6$\cite{Gong2017}. Although V$_5$S$_8$ bulk is an Ising antiferromagnet, the weak spin anisotropy may push thin layers toward a weakly anisotropic Heisenberg spin system. Further study on thinner flakes and with improved sample growth techniques is needed to unveil the origin.

\subsection{Magnetic phase diagram}

\begin{figure}[htbp]
\includegraphics[width=1\textwidth]{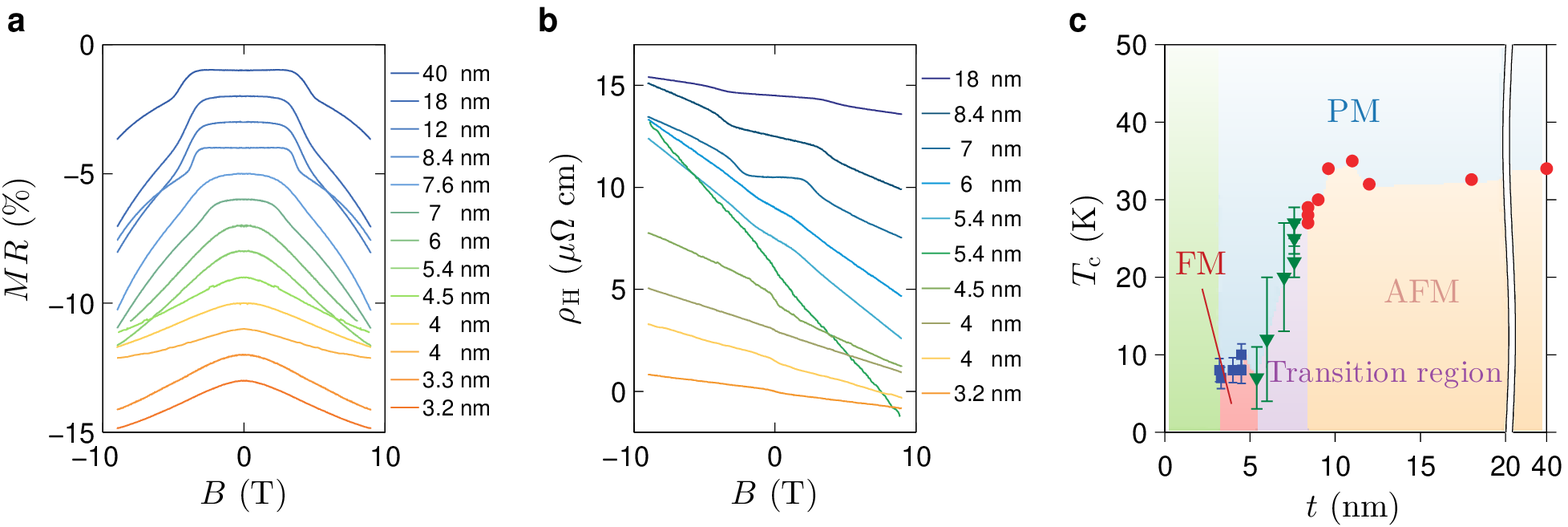}
\caption{Thickness dependence of magnetotransport and the phase diagram. Evolution of (a) MR and (b) $\rho_\text{H}$ with thickness at 1.5 K. For clarity, only a part of the samples are shown. More data can be found in the Supplemental Material. (c) Critical temperature $T_\text{c}$ - thickness $t$ phase diagram.}
\label{fig5}
\end{figure}

To give an overall picture of the evolution of transport properties with thickness, the low temperature MR and Hall resistivity for samples of a series of thicknesses are plotted in \rfig{fig5}a and b. For thicker samples, MR shows a plateau in low fields and a sudden drop at the critical field $H_\text{c}$ of the SF transition, followed by a quadratic field dependence. Correspondingly, the Hall displays a jump at $H_\text{c}$. As the thickness goes below 8.4 nm, the SF transition fades out. In particular, the discontinuity in MR and Hall is smeared out and $H_\text{c}$ is reduced. At last, the MR is quadratic except for a tendency toward a sublinear dependence in high fields. The Hall is close to linear, suggesting a dominant PM behavior. With further decreasing thickness, an FM order develops, evidenced by a nonlinear AHE. From these observations, we are able to plot the magnetic phase diagram of the material as a function of thickness, illustrated in \rfig{fig5}c.

An intriguing thickness-induced magnetic phase transition is observed in V$_5$S$_8$. In contrast, the AFM ordering in FePS$_3$ has been found to persist down to monolayer\cite{Lee2016,Wang2016}. The phase transition we observed stems from competing AFM and FM interactions, which are manifested in the spin-glass-like state between the AFM and FM phases. The existence of an FM interaction is not surprising, as it can be inferred from the AFM ordering in the bulk material, shown in \rfig{fig1}c. Along the $x$ direction, the magnetic moments align in an alternating fashion, parallel and antiparallel. The parallel alignment suggests an FM interaction, which is in fact implied by a positive Currie-Weiss temperature, reported in previous studies \cite{Vries1973,Nozaki1978} and observed in the current work (see the Supplemental Material), as well. Thus, the reduction of thickness changes the balance between FM and AFM interactions, leading to the emergence of an FM ordering. Various mechanisms can potentially contribute to the dependence of the magnetic interactions on thickness, e.g., modification of the energy band due to quantum confinement, or change of the interlayer coupling. More work is required to identify and understand the effect.

\section{Conclusion}

Our findings demonstrate the AHE as a useful tool for magnetization characterization of 2D crystals and reveal an intriguing magnetic phase transition induced by reduced dimensionality in V$_5$S$_8$. Besides potential applications in spintronics, such materials could provide a new playground for studying magnetism, as not only the low dimensionality can qualitatively change the phenomenon, but the tunability of 2D materials enables studies over parameter spaces that are difficult or even impossible to access. In our thinnest samples, we have been able to see appearance of the gate tunability. Furthermore, compared with these newly found magnetic 2D materials, such as CrI$_3$, FePS$_3$ and Cr$_2$Ge$_2$Te$_6$\cite{Lee2016,Wang2016,Gong2017,Huang2017}, V$_5$S$_8$ is unique in that it is very stable in air and conductive. It may be an ideal platform for studying itinerant-electron magnetism in 2D.

\begin{acknowledgements}
This work was supported by The National Key Research and Development Program of China (Grants No. 2016YFA0300600, No. 2013CBA01603, and No. 2016YFA0300802) and NSFC (Project No. 11574005, No. 11222436, No. 11234001, No. 51290272, and No. 51472008).
\end{acknowledgements}



\begin{thebibliography}{36}%
\makeatletter
\providecommand \@ifxundefined [1]{%
 \@ifx{#1\undefined}
}%
\providecommand \@ifnum [1]{%
 \ifnum #1\expandafter \@firstoftwo
 \else \expandafter \@secondoftwo
 \fi
}%
\providecommand \@ifx [1]{%
 \ifx #1\expandafter \@firstoftwo
 \else \expandafter \@secondoftwo
 \fi
}%
\providecommand \natexlab [1]{#1}%
\providecommand \enquote  [1]{``#1''}%
\providecommand \bibnamefont  [1]{#1}%
\providecommand \bibfnamefont [1]{#1}%
\providecommand \citenamefont [1]{#1}%
\providecommand \href@noop [0]{\@secondoftwo}%
\providecommand \href [0]{\begingroup \@sanitize@url \@href}%
\providecommand \@href[1]{\@@startlink{#1}\@@href}%
\providecommand \@@href[1]{\endgroup#1\@@endlink}%
\providecommand \@sanitize@url [0]{\catcode `\\12\catcode `\$12\catcode
  `\&12\catcode `\#12\catcode `\^12\catcode `\_12\catcode `\%12\relax}%
\providecommand \@@startlink[1]{}%
\providecommand \@@endlink[0]{}%
\providecommand \url  [0]{\begingroup\@sanitize@url \@url }%
\providecommand \@url [1]{\endgroup\@href {#1}{\urlprefix }}%
\providecommand \urlprefix  [0]{URL }%
\providecommand \Eprint [0]{\href }%
\providecommand \doibase [0]{http://dx.doi.org/}%
\providecommand \selectlanguage [0]{\@gobble}%
\providecommand \bibinfo  [0]{\@secondoftwo}%
\providecommand \bibfield  [0]{\@secondoftwo}%
\providecommand \translation [1]{[#1]}%
\providecommand \BibitemOpen [0]{}%
\providecommand \bibitemStop [0]{}%
\providecommand \bibitemNoStop [0]{.\EOS\space}%
\providecommand \EOS [0]{\spacefactor3000\relax}%
\providecommand \BibitemShut  [1]{\csname bibitem#1\endcsname}%
\let\auto@bib@innerbib\@empty
\bibitem [{\citenamefont {Geim}\ and\ \citenamefont
  {Grigorieva}(2013)}]{Geim2013}%
  \BibitemOpen
  \bibfield  {author} {\bibinfo {author} {\bibfnamefont {A.~K.}\ \bibnamefont
  {Geim}}\ and\ \bibinfo {author} {\bibfnamefont {I.~V.}\ \bibnamefont
  {Grigorieva}},\ }\href {http://dx.doi.org/10.1038/nature12385} {\bibfield
  {journal} {\bibinfo  {journal} {Nature}\ }\textbf {\bibinfo {volume} {499}},\
  \bibinfo {pages} {419} (\bibinfo {year} {2013})}\BibitemShut {NoStop}%
\bibitem [{\citenamefont {Butler}\ \emph {et~al.}(2013)\citenamefont {Butler},
  \citenamefont {Hollen}, \citenamefont {Cao}, \citenamefont {Cui},
  \citenamefont {Gupta}, \citenamefont {Guti鑼卹rez}, \citenamefont {Heinz},
  \citenamefont {Hong}, \citenamefont {Huang}, \citenamefont {Ismach},
  \citenamefont {Johnston-Halperin}, \citenamefont {Kuno}, \citenamefont
  {Plashnitsa}, \citenamefont {Robinson}, \citenamefont {Ruoff}, \citenamefont
  {Salahuddin}, \citenamefont {Shan}, \citenamefont {Shi}, \citenamefont
  {Spencer}, \citenamefont {Terrones}, \citenamefont {Windl},\ and\
  \citenamefont {Goldberger}}]{Butler2013}%
  \BibitemOpen
  \bibfield  {author} {\bibinfo {author} {\bibfnamefont {S.~Z.}\ \bibnamefont
  {Butler}}, \bibinfo {author} {\bibfnamefont {S.~M.}\ \bibnamefont {Hollen}},
  \bibinfo {author} {\bibfnamefont {L.}~\bibnamefont {Cao}}, \bibinfo {author}
  {\bibfnamefont {Y.}~\bibnamefont {Cui}}, \bibinfo {author} {\bibfnamefont
  {J.~A.}\ \bibnamefont {Gupta}}, \bibinfo {author} {\bibfnamefont {H.~R.}\
  \bibnamefont {Guti鑼卹rez}}, \bibinfo {author} {\bibfnamefont {T.~F.}\
  \bibnamefont {Heinz}}, \bibinfo {author} {\bibfnamefont {S.~S.}\ \bibnamefont
  {Hong}}, \bibinfo {author} {\bibfnamefont {J.}~\bibnamefont {Huang}},
  \bibinfo {author} {\bibfnamefont {A.~F.}\ \bibnamefont {Ismach}}, \bibinfo
  {author} {\bibfnamefont {E.}~\bibnamefont {Johnston-Halperin}}, \bibinfo
  {author} {\bibfnamefont {M.}~\bibnamefont {Kuno}}, \bibinfo {author}
  {\bibfnamefont {V.~V.}\ \bibnamefont {Plashnitsa}}, \bibinfo {author}
  {\bibfnamefont {R.~D.}\ \bibnamefont {Robinson}}, \bibinfo {author}
  {\bibfnamefont {R.~S.}\ \bibnamefont {Ruoff}}, \bibinfo {author}
  {\bibfnamefont {S.}~\bibnamefont {Salahuddin}}, \bibinfo {author}
  {\bibfnamefont {J.}~\bibnamefont {Shan}}, \bibinfo {author} {\bibfnamefont
  {L.}~\bibnamefont {Shi}}, \bibinfo {author} {\bibfnamefont {M.~G.}\
  \bibnamefont {Spencer}}, \bibinfo {author} {\bibfnamefont {M.}~\bibnamefont
  {Terrones}}, \bibinfo {author} {\bibfnamefont {W.}~\bibnamefont {Windl}}, \
  and\ \bibinfo {author} {\bibfnamefont {J.~E.}\ \bibnamefont {Goldberger}},\
  }\href {\doibase 10.1021/nn400280c} {\bibfield  {journal} {\bibinfo
  {journal} {ACS Nano}\ }\textbf {\bibinfo {volume} {7}},\ \bibinfo {pages}
  {2898} (\bibinfo {year} {2013})}\BibitemShut {NoStop}%
\bibitem [{\citenamefont {Li}\ \emph {et~al.}(2014)\citenamefont {Li},
  \citenamefont {Yu}, \citenamefont {Ye}, \citenamefont {Ge}, \citenamefont
  {Ou}, \citenamefont {Wu}, \citenamefont {Feng}, \citenamefont {Chen},\ and\
  \citenamefont {Zhang}}]{Li2014b}%
  \BibitemOpen
  \bibfield  {author} {\bibinfo {author} {\bibfnamefont {L.}~\bibnamefont
  {Li}}, \bibinfo {author} {\bibfnamefont {Y.}~\bibnamefont {Yu}}, \bibinfo
  {author} {\bibfnamefont {G.~J.}\ \bibnamefont {Ye}}, \bibinfo {author}
  {\bibfnamefont {Q.}~\bibnamefont {Ge}}, \bibinfo {author} {\bibfnamefont
  {X.}~\bibnamefont {Ou}}, \bibinfo {author} {\bibfnamefont {H.}~\bibnamefont
  {Wu}}, \bibinfo {author} {\bibfnamefont {D.}~\bibnamefont {Feng}}, \bibinfo
  {author} {\bibfnamefont {X.~H.}\ \bibnamefont {Chen}}, \ and\ \bibinfo
  {author} {\bibfnamefont {Y.}~\bibnamefont {Zhang}},\ }\href
  {http://dx.doi.org/10.1038/nnano.2014.35} {\bibfield  {journal} {\bibinfo
  {journal} {Nat Nano}\ }\textbf {\bibinfo {volume} {9}},\ \bibinfo {pages}
  {372} (\bibinfo {year} {2014})}\BibitemShut {NoStop}%
\bibitem [{\citenamefont {Zhang}\ \emph {et~al.}(2007)\citenamefont {Zhang},
  \citenamefont {Soon}, \citenamefont {Loh}, \citenamefont {Yin}, \citenamefont
  {Ding}, \citenamefont {Sullivian},\ and\ \citenamefont {Wu}}]{Zhang2007}%
  \BibitemOpen
  \bibfield  {author} {\bibinfo {author} {\bibfnamefont {J.}~\bibnamefont
  {Zhang}}, \bibinfo {author} {\bibfnamefont {J.~M.}\ \bibnamefont {Soon}},
  \bibinfo {author} {\bibfnamefont {K.~P.}\ \bibnamefont {Loh}}, \bibinfo
  {author} {\bibfnamefont {J.~H.}\ \bibnamefont {Yin}}, \bibinfo {author}
  {\bibfnamefont {J.}~\bibnamefont {Ding}}, \bibinfo {author} {\bibfnamefont
  {M.~B.}\ \bibnamefont {Sullivian}}, \ and\ \bibinfo {author} {\bibfnamefont
  {P.}~\bibnamefont {Wu}},\ }\href {\doibase 10.1021/nl071016r} {\bibfield
  {journal} {\bibinfo  {journal} {Nano Lett.}\ }\textbf {\bibinfo {volume}
  {7}},\ \bibinfo {pages} {2370} (\bibinfo {year} {2007})}\BibitemShut
  {NoStop}%
\bibitem [{\citenamefont {Wang}\ \emph {et~al.}(2009)\citenamefont {Wang},
  \citenamefont {Huang}, \citenamefont {Song}, \citenamefont {Zhang},
  \citenamefont {Ma}, \citenamefont {Liang},\ and\ \citenamefont
  {Chen}}]{Wang2009}%
  \BibitemOpen
  \bibfield  {author} {\bibinfo {author} {\bibfnamefont {Y.}~\bibnamefont
  {Wang}}, \bibinfo {author} {\bibfnamefont {Y.}~\bibnamefont {Huang}},
  \bibinfo {author} {\bibfnamefont {Y.}~\bibnamefont {Song}}, \bibinfo {author}
  {\bibfnamefont {X.}~\bibnamefont {Zhang}}, \bibinfo {author} {\bibfnamefont
  {Y.}~\bibnamefont {Ma}}, \bibinfo {author} {\bibfnamefont {J.}~\bibnamefont
  {Liang}}, \ and\ \bibinfo {author} {\bibfnamefont {Y.}~\bibnamefont {Chen}},\
  }\href {\doibase 10.1021/nl802810g} {\bibfield  {journal} {\bibinfo
  {journal} {Nano Lett.}\ }\textbf {\bibinfo {volume} {9}},\ \bibinfo {pages}
  {220} (\bibinfo {year} {2009})}\BibitemShut {NoStop}%
\bibitem [{\citenamefont {Tongay}\ \emph {et~al.}(2012)\citenamefont {Tongay},
  \citenamefont {Varnoosfaderani}, \citenamefont {Appleton}, \citenamefont
  {Wu},\ and\ \citenamefont {Hebard}}]{Tongay2012}%
  \BibitemOpen
  \bibfield  {author} {\bibinfo {author} {\bibfnamefont {S.}~\bibnamefont
  {Tongay}}, \bibinfo {author} {\bibfnamefont {S.~S.}\ \bibnamefont
  {Varnoosfaderani}}, \bibinfo {author} {\bibfnamefont {B.~R.}\ \bibnamefont
  {Appleton}}, \bibinfo {author} {\bibfnamefont {J.}~\bibnamefont {Wu}}, \ and\
  \bibinfo {author} {\bibfnamefont {A.~F.}\ \bibnamefont {Hebard}},\ }\href
  {\doibase http://dx.doi.org/10.1063/1.4753797} {\bibfield  {journal}
  {\bibinfo  {journal} {Appl. Phys. Lett.}\ }\textbf {\bibinfo {volume}
  {101}},\ \bibinfo {pages} {123105} (\bibinfo {year} {2012})}\BibitemShut
  {NoStop}%
\bibitem [{\citenamefont {Nair}\ \emph {et~al.}(2012)\citenamefont {Nair},
  \citenamefont {Sepioni}, \citenamefont {Tsai}, \citenamefont {Lehtinen},
  \citenamefont {Keinonen}, \citenamefont {Krasheninnikov}, \citenamefont
  {Thomson}, \citenamefont {Geim},\ and\ \citenamefont
  {Grigorieva}}]{Nair2012}%
  \BibitemOpen
  \bibfield  {author} {\bibinfo {author} {\bibfnamefont {R.~R.}\ \bibnamefont
  {Nair}}, \bibinfo {author} {\bibfnamefont {M.}~\bibnamefont {Sepioni}},
  \bibinfo {author} {\bibfnamefont {I.-L.}\ \bibnamefont {Tsai}}, \bibinfo
  {author} {\bibfnamefont {O.}~\bibnamefont {Lehtinen}}, \bibinfo {author}
  {\bibfnamefont {J.}~\bibnamefont {Keinonen}}, \bibinfo {author}
  {\bibfnamefont {A.~V.}\ \bibnamefont {Krasheninnikov}}, \bibinfo {author}
  {\bibfnamefont {T.}~\bibnamefont {Thomson}}, \bibinfo {author} {\bibfnamefont
  {A.~K.}\ \bibnamefont {Geim}}, \ and\ \bibinfo {author} {\bibfnamefont
  {I.~V.}\ \bibnamefont {Grigorieva}},\ }\href
  {http://dx.doi.org/10.1038/nphys2183} {\bibfield  {journal} {\bibinfo
  {journal} {Nat Phys}\ }\textbf {\bibinfo {volume} {8}},\ \bibinfo {pages}
  {199} (\bibinfo {year} {2012})}\BibitemShut {NoStop}%
\bibitem [{\citenamefont {Nair}\ \emph {et~al.}(2013)\citenamefont {Nair},
  \citenamefont {Tsai}, \citenamefont {Sepioni}, \citenamefont {Lehtinen},
  \citenamefont {Keinonen}, \citenamefont {Krasheninnikov}, \citenamefont
  {Castro~Neto}, \citenamefont {Katsnelson}, \citenamefont {Geim},\ and\
  \citenamefont {Grigorieva}}]{Nair2013}%
  \BibitemOpen
  \bibfield  {author} {\bibinfo {author} {\bibfnamefont {R.}~\bibnamefont
  {Nair}}, \bibinfo {author} {\bibfnamefont {I.-L.}\ \bibnamefont {Tsai}},
  \bibinfo {author} {\bibfnamefont {M.}~\bibnamefont {Sepioni}}, \bibinfo
  {author} {\bibfnamefont {O.}~\bibnamefont {Lehtinen}}, \bibinfo {author}
  {\bibfnamefont {J.}~\bibnamefont {Keinonen}}, \bibinfo {author}
  {\bibfnamefont {A.}~\bibnamefont {Krasheninnikov}}, \bibinfo {author}
  {\bibfnamefont {A.}~\bibnamefont {Castro~Neto}}, \bibinfo {author}
  {\bibfnamefont {M.}~\bibnamefont {Katsnelson}}, \bibinfo {author}
  {\bibfnamefont {A.}~\bibnamefont {Geim}}, \ and\ \bibinfo {author}
  {\bibfnamefont {I.}~\bibnamefont {Grigorieva}},\ }\href
  {http://dx.doi.org/10.1038/ncomms3010} {\bibfield  {journal} {\bibinfo
  {journal} {Nat Commun}\ }\textbf {\bibinfo {volume} {4}},\ \bibinfo {pages}
  {2010} (\bibinfo {year} {2013})}\BibitemShut {NoStop}%
\bibitem [{\citenamefont {Han}\ \emph {et~al.}(2014)\citenamefont {Han},
  \citenamefont {Kawakami}, \citenamefont {Gmitra},\ and\ \citenamefont
  {Fabian}}]{Han2014}%
  \BibitemOpen
  \bibfield  {author} {\bibinfo {author} {\bibfnamefont {W.}~\bibnamefont
  {Han}}, \bibinfo {author} {\bibfnamefont {R.~K.}\ \bibnamefont {Kawakami}},
  \bibinfo {author} {\bibfnamefont {M.}~\bibnamefont {Gmitra}}, \ and\ \bibinfo
  {author} {\bibfnamefont {J.}~\bibnamefont {Fabian}},\ }\href {\doibase
  10.1038/nnano.2014.214} {\bibfield  {journal} {\bibinfo  {journal} {Nat.
  Nanotechnol.}\ }\textbf {\bibinfo {volume} {9}},\ \bibinfo {pages} {794}
  (\bibinfo {year} {2014})}\BibitemShut {NoStop}%
\bibitem [{\citenamefont {Gonz谩lez-Herrero}\ \emph {et~al.}(2016)\citenamefont
  {Gonz谩lez-Herrero}, \citenamefont {G贸mez-Rodr铆guez}, \citenamefont
  {Mallet}, \citenamefont {Moaied}, \citenamefont {Palacios}, \citenamefont
  {Salgado}, \citenamefont {Ugeda}, \citenamefont {Veuillen}, \citenamefont
  {Yndurain},\ and\ \citenamefont {Brihuega}}]{Gonzalez-Herrero2016}%
  \BibitemOpen
  \bibfield  {author} {\bibinfo {author} {\bibfnamefont {H.}~\bibnamefont
  {Gonz谩lez-Herrero}}, \bibinfo {author} {\bibfnamefont {J.~M.}\ \bibnamefont
  {G贸mez-Rodr铆guez}}, \bibinfo {author} {\bibfnamefont {P.}~\bibnamefont
  {Mallet}}, \bibinfo {author} {\bibfnamefont {M.}~\bibnamefont {Moaied}},
  \bibinfo {author} {\bibfnamefont {J.~J.}\ \bibnamefont {Palacios}}, \bibinfo
  {author} {\bibfnamefont {C.}~\bibnamefont {Salgado}}, \bibinfo {author}
  {\bibfnamefont {M.~M.}\ \bibnamefont {Ugeda}}, \bibinfo {author}
  {\bibfnamefont {J.-Y.}\ \bibnamefont {Veuillen}}, \bibinfo {author}
  {\bibfnamefont {F.}~\bibnamefont {Yndurain}}, \ and\ \bibinfo {author}
  {\bibfnamefont {I.}~\bibnamefont {Brihuega}},\ }\href
  {http://science.sciencemag.org/content/352/6284/437.abstract} {\bibfield
  {journal} {\bibinfo  {journal} {Science}\ }\textbf {\bibinfo {volume}
  {352}},\ \bibinfo {pages} {437} (\bibinfo {year} {2016})}\BibitemShut
  {NoStop}%
\bibitem [{\citenamefont {Du}\ \emph {et~al.}(2016)\citenamefont {Du},
  \citenamefont {Wang}, \citenamefont {Liu}, \citenamefont {Hu}, \citenamefont
  {Utama}, \citenamefont {Gan}, \citenamefont {Xiong},\ and\ \citenamefont
  {Kloc}}]{Du2016}%
  \BibitemOpen
  \bibfield  {author} {\bibinfo {author} {\bibfnamefont {K.-z.}\ \bibnamefont
  {Du}}, \bibinfo {author} {\bibfnamefont {X.-z.}\ \bibnamefont {Wang}},
  \bibinfo {author} {\bibfnamefont {Y.}~\bibnamefont {Liu}}, \bibinfo {author}
  {\bibfnamefont {P.}~\bibnamefont {Hu}}, \bibinfo {author} {\bibfnamefont
  {M.~I.~B.}\ \bibnamefont {Utama}}, \bibinfo {author} {\bibfnamefont {C.~K.}\
  \bibnamefont {Gan}}, \bibinfo {author} {\bibfnamefont {Q.}~\bibnamefont
  {Xiong}}, \ and\ \bibinfo {author} {\bibfnamefont {C.}~\bibnamefont {Kloc}},\
  }\href {\doibase 10.1021/acsnano.5b05927} {\bibfield  {journal} {\bibinfo
  {journal} {ACS Nano}\ }\textbf {\bibinfo {volume} {10}},\ \bibinfo {pages}
  {1738} (\bibinfo {year} {2016})}\BibitemShut {NoStop}%
\bibitem [{\citenamefont {Park}(2016)}]{Park2016}%
  \BibitemOpen
  \bibfield  {author} {\bibinfo {author} {\bibfnamefont {J.-G.}\ \bibnamefont
  {Park}},\ }\href {http://stacks.iop.org/0953-8984/28/i=30/a=301001}
  {\bibfield  {journal} {\bibinfo  {journal} {J. Phys.: Condens. Matter}\
  }\textbf {\bibinfo {volume} {28}},\ \bibinfo {pages} {301001} (\bibinfo
  {year} {2016})}\BibitemShut {NoStop}%
\bibitem [{\citenamefont {Lee}\ \emph {et~al.}(2016)\citenamefont {Lee},
  \citenamefont {Lee}, \citenamefont {Ryoo}, \citenamefont {Kang},
  \citenamefont {Kim}, \citenamefont {Kim}, \citenamefont {Park}, \citenamefont
  {Park},\ and\ \citenamefont {Cheong}}]{Lee2016}%
  \BibitemOpen
  \bibfield  {author} {\bibinfo {author} {\bibfnamefont {J.-U.}\ \bibnamefont
  {Lee}}, \bibinfo {author} {\bibfnamefont {S.}~\bibnamefont {Lee}}, \bibinfo
  {author} {\bibfnamefont {J.~H.}\ \bibnamefont {Ryoo}}, \bibinfo {author}
  {\bibfnamefont {S.}~\bibnamefont {Kang}}, \bibinfo {author} {\bibfnamefont
  {T.~Y.}\ \bibnamefont {Kim}}, \bibinfo {author} {\bibfnamefont
  {P.}~\bibnamefont {Kim}}, \bibinfo {author} {\bibfnamefont {C.-H.}\
  \bibnamefont {Park}}, \bibinfo {author} {\bibfnamefont {J.-G.}\ \bibnamefont
  {Park}}, \ and\ \bibinfo {author} {\bibfnamefont {H.}~\bibnamefont
  {Cheong}},\ }\href {\doibase 10.1021/acs.nanolett.6b03052} {\bibfield
  {journal} {\bibinfo  {journal} {Nano Lett.}\ }\textbf {\bibinfo {volume}
  {16}},\ \bibinfo {pages} {7433} (\bibinfo {year} {2016})}\BibitemShut {NoStop}%
\bibitem [{\citenamefont {Wang}\ \emph {et~al.}(2016)\citenamefont {Wang},
  \citenamefont {Du}, \citenamefont {Liu}, \citenamefont {Hu}, \citenamefont
  {Zhang}, \citenamefont {Zhang}, \citenamefont {Owen}, \citenamefont {Lu},
  \citenamefont {Gan}, \citenamefont {Sengupta}, \citenamefont {Kloc},\ and\
  \citenamefont {Xiong}}]{Wang2016}%
  \BibitemOpen
  \bibfield  {author} {\bibinfo {author} {\bibfnamefont {X.}~\bibnamefont
  {Wang}}, \bibinfo {author} {\bibfnamefont {K.}~\bibnamefont {Du}}, \bibinfo
  {author} {\bibfnamefont {Y.~Y.~F.}\ \bibnamefont {Liu}}, \bibinfo {author}
  {\bibfnamefont {P.}~\bibnamefont {Hu}}, \bibinfo {author} {\bibfnamefont
  {J.}~\bibnamefont {Zhang}}, \bibinfo {author} {\bibfnamefont
  {Q.}~\bibnamefont {Zhang}}, \bibinfo {author} {\bibfnamefont {M.~H.~S.}\
  \bibnamefont {Owen}}, \bibinfo {author} {\bibfnamefont {X.}~\bibnamefont
  {Lu}}, \bibinfo {author} {\bibfnamefont {C.~K.}\ \bibnamefont {Gan}},
  \bibinfo {author} {\bibfnamefont {P.}~\bibnamefont {Sengupta}}, \bibinfo
  {author} {\bibfnamefont {C.}~\bibnamefont {Kloc}}, \ and\ \bibinfo {author}
  {\bibfnamefont {Q.}~\bibnamefont {Xiong}},\ }\href
  {http://stacks.iop.org/2053-1583/3/i=3/a=031009} {\bibfield  {journal}
  {\bibinfo  {journal} {2D Materials}\ }\textbf {\bibinfo {volume} {3}},\
  \bibinfo {pages} {031009} (\bibinfo {year} {2016})}\BibitemShut {NoStop}%
\bibitem [{\citenamefont {Gong}\ \emph {et~al.}(2017)\citenamefont {Gong},
  \citenamefont {Li}, \citenamefont {Li}, \citenamefont {Ji}, \citenamefont
  {Stern}, \citenamefont {Xia}, \citenamefont {Cao}, \citenamefont {Bao},
  \citenamefont {Wang}, \citenamefont {Wang}, \citenamefont {Qiu},
  \citenamefont {Cava}, \citenamefont {Louie}, \citenamefont {Xia},\ and\
  \citenamefont {Zhang}}]{Gong2017}%
  \BibitemOpen
  \bibfield  {author} {\bibinfo {author} {\bibfnamefont {C.}~\bibnamefont
  {Gong}}, \bibinfo {author} {\bibfnamefont {L.}~\bibnamefont {Li}}, \bibinfo
  {author} {\bibfnamefont {Z.}~\bibnamefont {Li}}, \bibinfo {author}
  {\bibfnamefont {H.}~\bibnamefont {Ji}}, \bibinfo {author} {\bibfnamefont
  {A.}~\bibnamefont {Stern}}, \bibinfo {author} {\bibfnamefont
  {Y.}~\bibnamefont {Xia}}, \bibinfo {author} {\bibfnamefont {T.}~\bibnamefont
  {Cao}}, \bibinfo {author} {\bibfnamefont {W.}~\bibnamefont {Bao}}, \bibinfo
  {author} {\bibfnamefont {C.}~\bibnamefont {Wang}}, \bibinfo {author}
  {\bibfnamefont {Y.}~\bibnamefont {Wang}}, \bibinfo {author} {\bibfnamefont
  {Z.~Q.}\ \bibnamefont {Qiu}}, \bibinfo {author} {\bibfnamefont {R.~J.}\
  \bibnamefont {Cava}}, \bibinfo {author} {\bibfnamefont {S.~G.}\ \bibnamefont
  {Louie}}, \bibinfo {author} {\bibfnamefont {J.}~\bibnamefont {Xia}}, \ and\
  \bibinfo {author} {\bibfnamefont {X.}~\bibnamefont {Zhang}},\ }\href@noop {}
  \ \Eprint
  {http://arxiv.org/abs/1703.05753} {arXiv:1703.05753} \BibitemShut {NoStop}%
\bibitem [{\citenamefont {Huang}\ \emph {et~al.}(2017)\citenamefont {Huang},
  \citenamefont {Clark}, \citenamefont {Navarro-Moratalla}, \citenamefont
  {Klein}, \citenamefont {Cheng}, \citenamefont {Seyler}, \citenamefont
  {Zhong}, \citenamefont {Schmidgall}, \citenamefont {McGuire}, \citenamefont
  {Cobden}, \citenamefont {Yao}, \citenamefont {Xiao}, \citenamefont
  {Jarillo-Herrero},\ and\ \citenamefont {Xu}}]{Huang2017}%
  \BibitemOpen
  \bibfield  {author} {\bibinfo {author} {\bibfnamefont {B.}~\bibnamefont
  {Huang}}, \bibinfo {author} {\bibfnamefont {G.}~\bibnamefont {Clark}},
  \bibinfo {author} {\bibfnamefont {E.}~\bibnamefont {Navarro-Moratalla}},
  \bibinfo {author} {\bibfnamefont {D.~R.}\ \bibnamefont {Klein}}, \bibinfo
  {author} {\bibfnamefont {R.}~\bibnamefont {Cheng}}, \bibinfo {author}
  {\bibfnamefont {K.~L.}\ \bibnamefont {Seyler}}, \bibinfo {author}
  {\bibfnamefont {D.}~\bibnamefont {Zhong}}, \bibinfo {author} {\bibfnamefont
  {E.}~\bibnamefont {Schmidgall}}, \bibinfo {author} {\bibfnamefont {M.~A.}\
  \bibnamefont {McGuire}}, \bibinfo {author} {\bibfnamefont {D.~H.}\
  \bibnamefont {Cobden}}, \bibinfo {author} {\bibfnamefont {W.}~\bibnamefont
  {Yao}}, \bibinfo {author} {\bibfnamefont {D.}~\bibnamefont {Xiao}}, \bibinfo
  {author} {\bibfnamefont {P.}~\bibnamefont {Jarillo-Herrero}}, \ and\ \bibinfo
  {author} {\bibfnamefont {X.}~\bibnamefont {Xu}},\ }\href@noop {} \ \Eprint
  {http://arxiv.org/abs/1703.05892} {arXiv:1703.05892} \BibitemShut {NoStop}%
\bibitem [{\citenamefont {Kawada}\ \emph {et~al.}(1975)\citenamefont {Kawada},
  \citenamefont {Nakano-Onoda}, \citenamefont {Ishii}, \citenamefont {Saeki},\
  and\ \citenamefont {Nakahira}}]{Kawada1975}%
  \BibitemOpen
  \bibfield  {author} {\bibinfo {author} {\bibfnamefont {I.}~\bibnamefont
  {Kawada}}, \bibinfo {author} {\bibfnamefont {M.}~\bibnamefont
  {Nakano-Onoda}}, \bibinfo {author} {\bibfnamefont {M.}~\bibnamefont {Ishii}},
  \bibinfo {author} {\bibfnamefont {M.}~\bibnamefont {Saeki}}, \ and\ \bibinfo
  {author} {\bibfnamefont {M.}~\bibnamefont {Nakahira}},\ }\href {\doibase
  http://dx.doi.org/10.1016/0022-4596(75)90209-1} {\bibfield  {journal}
  {\bibinfo  {journal} {J. Solid State Chem.}\ }\textbf {\bibinfo {volume}
  {15}},\ \bibinfo {pages} {246} (\bibinfo {year} {1975})}\BibitemShut
  {NoStop}%
\bibitem [{\citenamefont {Vries}\ and\ \citenamefont {Haas}(1973)}]{Vries1973}%
  \BibitemOpen
  \bibfield  {author} {\bibinfo {author} {\bibfnamefont {A.~D.}\ \bibnamefont
  {Vries}}\ and\ \bibinfo {author} {\bibfnamefont {C.}~\bibnamefont {Haas}},\
  }\href {\doibase http://dx.doi.org/10.1016/S0022-3697(73)80171-4} {\bibfield
  {journal} {\bibinfo  {journal} {J. Phys. Chem. Solids}\
  }\textbf {\bibinfo {volume} {34}},\ \bibinfo {pages} {651} (\bibinfo {year}
  {1973})}\BibitemShut {NoStop}%
\bibitem [{\citenamefont {Silbernagel}\ \emph {et~al.}(1975)\citenamefont
  {Silbernagel}, \citenamefont {Levy},\ and\ \citenamefont
  {Gamble}}]{Silbernagel1975}%
  \BibitemOpen
  \bibfield  {author} {\bibinfo {author} {\bibfnamefont {B.~G.}\ \bibnamefont
  {Silbernagel}}, \bibinfo {author} {\bibfnamefont {R.~B.}\ \bibnamefont
  {Levy}}, \ and\ \bibinfo {author} {\bibfnamefont {F.~R.}\ \bibnamefont
  {Gamble}},\ }\href {\doibase 10.1103/PhysRevB.11.4563} {\bibfield  {journal}
  {\bibinfo  {journal} {Phys. Rev. B}\ }\textbf {\bibinfo {volume} {11}},\
  \bibinfo {pages} {4563} (\bibinfo {year} {1975})}\BibitemShut {NoStop}%
\bibitem [{\citenamefont {Nozaki}\ and\ \citenamefont
  {Ishizawa}(1977)}]{Nozaki1977}%
  \BibitemOpen
  \bibfield  {author} {\bibinfo {author} {\bibfnamefont {H.}~\bibnamefont
  {Nozaki}}\ and\ \bibinfo {author} {\bibfnamefont {Y.}~\bibnamefont
  {Ishizawa}},\ }\href {\doibase 10.1016/0375-9601(77)90224-9} {\bibfield
  {journal} {\bibinfo  {journal} {Phys. Lett. A}\ }\textbf {\bibinfo
  {volume} {63}},\ \bibinfo {pages} {131} (\bibinfo {year} {1977})}\BibitemShut
  {NoStop}%
\bibitem [{\citenamefont {Nozaki}\ \emph {et~al.}(1978)\citenamefont {Nozaki},
  \citenamefont {Umehara}, \citenamefont {Ishizawa}, \citenamefont {Saeki},
  \citenamefont {Mizoguchi},\ and\ \citenamefont {Nakahira}}]{Nozaki1978}%
  \BibitemOpen
  \bibfield  {author} {\bibinfo {author} {\bibfnamefont {H.}~\bibnamefont
  {Nozaki}}, \bibinfo {author} {\bibfnamefont {M.}~\bibnamefont {Umehara}},
  \bibinfo {author} {\bibfnamefont {Y.}~\bibnamefont {Ishizawa}}, \bibinfo
  {author} {\bibfnamefont {M.}~\bibnamefont {Saeki}}, \bibinfo {author}
  {\bibfnamefont {T.}~\bibnamefont {Mizoguchi}}, \ and\ \bibinfo {author}
  {\bibfnamefont {M.}~\bibnamefont {Nakahira}},\ }\href
  {http://www.sciencedirect.com/science/article/pii/0022369778901440}
  {\bibfield  {journal} {\bibinfo  {journal} {J. Phys. Chem. Solids}\ }\textbf
  {\bibinfo {volume} {39}},\ \bibinfo {pages} {851} (\bibinfo {year}
  {1978})}\BibitemShut {NoStop}%
\bibitem [{\citenamefont {Kitaoka}\ and\ \citenamefont
  {Yasuoka}(1980)}]{Kitaoka1980}%
  \BibitemOpen
  \bibfield  {author} {\bibinfo {author} {\bibfnamefont {Y.}~\bibnamefont
  {Kitaoka}}\ and\ \bibinfo {author} {\bibfnamefont {H.}~\bibnamefont
  {Yasuoka}},\ }\href {\doibase 10.1143/JPSJ.48.1949} {\bibfield  {journal}
  {\bibinfo  {journal} {J. Phys. Soc. Jpn.}\ }\textbf {\bibinfo {volume}
  {48}},\ \bibinfo {pages} {1949} (\bibinfo {year} {1980})}\BibitemShut {NoStop}%
\bibitem [{\citenamefont {Funahashi}\ \emph {et~al.}(1981)\citenamefont
  {Funahashi}, \citenamefont {Nozaki},\ and\ \citenamefont
  {Kawada}}]{Funahashi1981}%
  \BibitemOpen
  \bibfield  {author} {\bibinfo {author} {\bibfnamefont {S.}~\bibnamefont
  {Funahashi}}, \bibinfo {author} {\bibfnamefont {H.}~\bibnamefont {Nozaki}}, \
  and\ \bibinfo {author} {\bibfnamefont {I.}~\bibnamefont {Kawada}},\ }\href
  {http://www.sciencedirect.com/science/article/pii/0022369781900640}
  {\bibfield  {journal} {\bibinfo  {journal} {J. Phys. Chem. Solids}\ }\textbf
  {\bibinfo {volume} {42}},\ \bibinfo {pages} {1009} (\bibinfo {year}
  {1981})}\BibitemShut {NoStop}%
\bibitem [{\citenamefont {Nakanishi}\ \emph {et~al.}(2000)\citenamefont
  {Nakanishi}, \citenamefont {Yoshimura}, \citenamefont {Kosuge}, \citenamefont
  {Goto}, \citenamefont {Fujii},\ and\ \citenamefont {Takada}}]{Nakanishi2000}%
  \BibitemOpen
  \bibfield  {author} {\bibinfo {author} {\bibfnamefont {M.}~\bibnamefont
  {Nakanishi}}, \bibinfo {author} {\bibfnamefont {K.}~\bibnamefont
  {Yoshimura}}, \bibinfo {author} {\bibfnamefont {K.}~\bibnamefont {Kosuge}},
  \bibinfo {author} {\bibfnamefont {T.}~\bibnamefont {Goto}}, \bibinfo {author}
  {\bibfnamefont {T.}~\bibnamefont {Fujii}}, \ and\ \bibinfo {author}
  {\bibfnamefont {J.}~\bibnamefont {Takada}},\ }\href {\doibase
  http://dx.doi.org/10.1016/S0304-8853(00)00509-6} {\bibfield  {journal}
  {\bibinfo  {journal} {J. Magn. Magn. Mater.}\ }\textbf
  {\bibinfo {volume} {221}},\ \bibinfo {pages} {301} (\bibinfo {year}
  {2000})}\BibitemShut {NoStop}%
\bibitem [{\citenamefont {Hardy}\ \emph {et~al.}(2016)\citenamefont {Hardy},
  \citenamefont {Yuan}, \citenamefont {Guo}, \citenamefont {Zhou},
  \citenamefont {Lou},\ and\ \citenamefont {Natelson}}]{Hardy2016}%
  \BibitemOpen
  \bibfield  {author} {\bibinfo {author} {\bibfnamefont {W.~J.}\ \bibnamefont
  {Hardy}}, \bibinfo {author} {\bibfnamefont {J.}~\bibnamefont {Yuan}},
  \bibinfo {author} {\bibfnamefont {H.}~\bibnamefont {Guo}}, \bibinfo {author}
  {\bibfnamefont {P.}~\bibnamefont {Zhou}}, \bibinfo {author} {\bibfnamefont
  {J.}~\bibnamefont {Lou}}, \ and\ \bibinfo {author} {\bibfnamefont
  {D.}~\bibnamefont {Natelson}},\ }\href {\doibase 10.1021/acsnano.6b01269}
  {\bibfield  {journal} {\bibinfo  {journal} {ACS Nano}\ }\textbf {\bibinfo
  {volume} {10}},\ \bibinfo {pages} {5941} (\bibinfo {year} {2016})}\BibitemShut {NoStop}%
\bibitem [{\citenamefont {Poddar}\ and\ \citenamefont
  {Rastogi}(2002)}]{Poddar2002}%
  \BibitemOpen
  \bibfield  {author} {\bibinfo {author} {\bibfnamefont {P.}~\bibnamefont
  {Poddar}}\ and\ \bibinfo {author} {\bibfnamefont {A.~K.}\ \bibnamefont
  {Rastogi}},\ }\href {http://stacks.iop.org/0953-8984/14/i=10/a=317}
  {\bibfield  {journal} {\bibinfo  {journal} {J. Phys.: Condens. Matter}\
  }\textbf {\bibinfo {volume} {14}},\ \bibinfo {pages} {2677} (\bibinfo {year}
  {2002})}\BibitemShut {NoStop}%
\bibitem [{\citenamefont {Oh}\ \emph {et~al.}(2014)\citenamefont {Oh},
  \citenamefont {Artyukhin}, \citenamefont {Yang}, \citenamefont {Zapf},
  \citenamefont {Kim}, \citenamefont {Vanderbilt},\ and\ \citenamefont
  {Cheong}}]{Oh2014}%
  \BibitemOpen
  \bibfield  {author} {\bibinfo {author} {\bibfnamefont {Y.~S.}\ \bibnamefont
  {Oh}}, \bibinfo {author} {\bibfnamefont {S.}~\bibnamefont {Artyukhin}},
  \bibinfo {author} {\bibfnamefont {J.~J.}\ \bibnamefont {Yang}}, \bibinfo
  {author} {\bibfnamefont {V.}~\bibnamefont {Zapf}}, \bibinfo {author}
  {\bibfnamefont {J.~W.}\ \bibnamefont {Kim}}, \bibinfo {author} {\bibfnamefont
  {D.}~\bibnamefont {Vanderbilt}}, \ and\ \bibinfo {author} {\bibfnamefont
  {S.-W.}\ \bibnamefont {Cheong}},\ }\href
  {http://dx.doi.org/10.1038/ncomms4201} {\bibfield  {journal} {\bibinfo
  {journal} {Nat Commun}\ }\textbf {\bibinfo {volume} {5}},\ \bibinfo {pages}
  {3201} (\bibinfo {year} {2014})}\BibitemShut {NoStop}%
\bibitem [{\citenamefont {Yamada}\ and\ \citenamefont
  {Takada}(1972)}]{Yamada1972}%
  \BibitemOpen
  \bibfield  {author} {\bibinfo {author} {\bibfnamefont {H.}~\bibnamefont
  {Yamada}}\ and\ \bibinfo {author} {\bibfnamefont {S.}~\bibnamefont
  {Takada}},\ }\href {\doibase 10.1143/ptp.48.1828} {\bibfield  {journal}
  {\bibinfo  {journal} {Prog. Theor. Phys.}\ }\textbf {\bibinfo {volume}
  {48}},\ \bibinfo {pages} {1828} (\bibinfo {year} {1972})}\BibitemShut
  {NoStop}%
\bibitem [{\citenamefont {Jin}\ \emph {et~al.}(2015)\citenamefont {Jin},
  \citenamefont {He}, \citenamefont {Zhang}, \citenamefont {Maruyama},
  \citenamefont {Yasui}, \citenamefont {Suchoski}, \citenamefont {Shin},
  \citenamefont {Jiang}, \citenamefont {Yu}, \citenamefont {Yuan},
  \citenamefont {Shan}, \citenamefont {Kusmartsev}, \citenamefont {Greene},\
  and\ \citenamefont {Takeuchi}}]{Jin2015}%
  \BibitemOpen
  \bibfield  {author} {\bibinfo {author} {\bibfnamefont {K.}~\bibnamefont
  {Jin}}, \bibinfo {author} {\bibfnamefont {G.}~\bibnamefont {He}}, \bibinfo
  {author} {\bibfnamefont {X.}~\bibnamefont {Zhang}}, \bibinfo {author}
  {\bibfnamefont {S.}~\bibnamefont {Maruyama}}, \bibinfo {author}
  {\bibfnamefont {S.}~\bibnamefont {Yasui}}, \bibinfo {author} {\bibfnamefont
  {R.}~\bibnamefont {Suchoski}}, \bibinfo {author} {\bibfnamefont
  {J.}~\bibnamefont {Shin}}, \bibinfo {author} {\bibfnamefont {Y.}~\bibnamefont
  {Jiang}}, \bibinfo {author} {\bibfnamefont {H.~S.}\ \bibnamefont {Yu}},
  \bibinfo {author} {\bibfnamefont {J.}~\bibnamefont {Yuan}}, \bibinfo {author}
  {\bibfnamefont {L.}~\bibnamefont {Shan}}, \bibinfo {author} {\bibfnamefont
  {F.~V.}\ \bibnamefont {Kusmartsev}}, \bibinfo {author} {\bibfnamefont
  {R.~L.}\ \bibnamefont {Greene}}, \ and\ \bibinfo {author} {\bibfnamefont
  {I.}~\bibnamefont {Takeuchi}},\ }\href {http://dx.doi.org/10.1038/ncomms8183}
  {\bibfield  {journal} {\bibinfo  {journal} {Nat. Commun.}\ }\textbf {\bibinfo
  {volume} {6}},\ \bibinfo {pages} {8183} (\bibinfo {year} {2015})}\BibitemShut
  {NoStop}%
\bibitem [{\citenamefont {Groot}\ and\ \citenamefont
  {Jongh}(1986)}]{Groot1986}%
  \BibitemOpen
  \bibfield  {author} {\bibinfo {author} {\bibfnamefont {H.~D.}\ \bibnamefont
  {Groot}}\ and\ \bibinfo {author} {\bibfnamefont {L.~D.}\ \bibnamefont
  {Jongh}},\ }\href {\doibase http://dx.doi.org/10.1016/0378-4363(86)90346-3}
  {\bibfield  {journal} {\bibinfo  {journal} {Physica B+C}\ }\textbf {\bibinfo
  {volume} {141}},\ \bibinfo {pages} {1 } (\bibinfo {year} {1986})}\BibitemShut
  {NoStop}%
\bibitem [{\citenamefont {Balamurugan}\ \emph {et~al.}(2014)\citenamefont
  {Balamurugan}, \citenamefont {Lee}, \citenamefont {Kim}, \citenamefont {Ok},
  \citenamefont {Jo}, \citenamefont {Song}, \citenamefont {Kim}, \citenamefont
  {Choi}, \citenamefont {Le},\ and\ \citenamefont {Park}}]{Balamurugan2014}%
  \BibitemOpen
  \bibfield  {author} {\bibinfo {author} {\bibfnamefont {K.}~\bibnamefont
  {Balamurugan}}, \bibinfo {author} {\bibfnamefont {S.-H.}\ \bibnamefont
  {Lee}}, \bibinfo {author} {\bibfnamefont {J.-S.}\ \bibnamefont {Kim}},
  \bibinfo {author} {\bibfnamefont {J.-M.}\ \bibnamefont {Ok}}, \bibinfo
  {author} {\bibfnamefont {Y.-J.}\ \bibnamefont {Jo}}, \bibinfo {author}
  {\bibfnamefont {Y.-M.}\ \bibnamefont {Song}}, \bibinfo {author}
  {\bibfnamefont {S.-A.}\ \bibnamefont {Kim}}, \bibinfo {author} {\bibfnamefont
  {E.~S.}\ \bibnamefont {Choi}}, \bibinfo {author} {\bibfnamefont {M.~D.}\
  \bibnamefont {Le}}, \ and\ \bibinfo {author} {\bibfnamefont {J.-G.}\
  \bibnamefont {Park}},\ }\href {\doibase 10.1103/PhysRevB.90.104412}
  {\bibfield  {journal} {\bibinfo  {journal} {Phys. Rev. B}\ }\textbf {\bibinfo
  {volume} {90}},\ \bibinfo {pages} {104412} (\bibinfo {year}
  {2014})}\BibitemShut {NoStop}%
\bibitem [{\citenamefont {Fert}\ and\ \citenamefont {Jaoul}(1972)}]{Fert1972}%
  \BibitemOpen
  \bibfield  {author} {\bibinfo {author} {\bibfnamefont {A.}~\bibnamefont
  {Fert}}\ and\ \bibinfo {author} {\bibfnamefont {O.}~\bibnamefont {Jaoul}},\
  }\href {\doibase 10.1103/PhysRevLett.28.303} {\bibfield  {journal} {\bibinfo
  {journal} {Phys. Rev. Lett.}\ }\textbf {\bibinfo {volume} {28}},\ \bibinfo
  {pages} {303} (\bibinfo {year} {1972})}\BibitemShut {NoStop}%
\bibitem [{\citenamefont {Nakatsuji}\ \emph {et~al.}(2015)\citenamefont
  {Nakatsuji}, \citenamefont {Kiyohara},\ and\ \citenamefont
  {Higo}}]{Nakatsuji2015}%
  \BibitemOpen
  \bibfield  {author} {\bibinfo {author} {\bibfnamefont {S.}~\bibnamefont
  {Nakatsuji}}, \bibinfo {author} {\bibfnamefont {N.}~\bibnamefont {Kiyohara}},
  \ and\ \bibinfo {author} {\bibfnamefont {T.}~\bibnamefont {Higo}},\ }\href
  {\doibase 10.1038/nature15723} {\bibfield  {journal} {\bibinfo  {journal}
  {Nature}\ }\textbf {\bibinfo {volume} {527}},\ \bibinfo {pages} {212}
  (\bibinfo {year} {2015})}\BibitemShut {NoStop}%
\bibitem [{\citenamefont {Luo}\ \emph {et~al.}(2015)\citenamefont {Luo},
  \citenamefont {Ronning}, \citenamefont {Wakeham}, \citenamefont {Lu},
  \citenamefont {Park}, \citenamefont {Xu},\ and\ \citenamefont
  {Thompson}}]{Luo2015}%
  \BibitemOpen
  \bibfield  {author} {\bibinfo {author} {\bibfnamefont {Y.}~\bibnamefont
  {Luo}}, \bibinfo {author} {\bibfnamefont {F.}~\bibnamefont {Ronning}},
  \bibinfo {author} {\bibfnamefont {N.}~\bibnamefont {Wakeham}}, \bibinfo
  {author} {\bibfnamefont {X.}~\bibnamefont {Lu}}, \bibinfo {author}
  {\bibfnamefont {T.}~\bibnamefont {Park}}, \bibinfo {author} {\bibfnamefont
  {Z.-A.}\ \bibnamefont {Xu}}, \ and\ \bibinfo {author} {\bibfnamefont {J.~D.}\
  \bibnamefont {Thompson}},\ }\href
  {http://www.pnas.org/content/112/44/13520.abstract} {\bibfield  {journal}
  {\bibinfo  {journal} {Proc. Natl. Acad. Sci.}\ }\textbf {\bibinfo {volume}
  {112}},\ \bibinfo {pages} {13520} (\bibinfo {year} {2015})}\BibitemShut
  {NoStop}%
\bibitem [{\citenamefont {Suzuki}\ \emph {et~al.}(2016)\citenamefont {Suzuki},
  \citenamefont {Chisnell}, \citenamefont {Devarakonda}, \citenamefont {Liu},
  \citenamefont {Feng}, \citenamefont {Xiao}, \citenamefont {Lynn},\ and\
  \citenamefont {Checkelsky}}]{Suzuki2016}%
  \BibitemOpen
  \bibfield  {author} {\bibinfo {author} {\bibfnamefont {T.}~\bibnamefont
  {Suzuki}}, \bibinfo {author} {\bibfnamefont {R.}~\bibnamefont {Chisnell}},
  \bibinfo {author} {\bibfnamefont {A.}~\bibnamefont {Devarakonda}}, \bibinfo
  {author} {\bibfnamefont {Y.-T.}\ \bibnamefont {Liu}}, \bibinfo {author}
  {\bibfnamefont {W.}~\bibnamefont {Feng}}, \bibinfo {author} {\bibfnamefont
  {D.}~\bibnamefont {Xiao}}, \bibinfo {author} {\bibfnamefont {J.~W.}\
  \bibnamefont {Lynn}}, \ and\ \bibinfo {author} {\bibfnamefont {J.~G.}\
  \bibnamefont {Checkelsky}},\ }\href {http://dx.doi.org/10.1038/nphys3831}
  {\bibfield  {journal} {\bibinfo  {journal} {Nat Phys}\ }\textbf {\bibinfo
  {volume} {12}},\ \bibinfo {pages} {1119} (\bibinfo {year}
  {2016})}\BibitemShut {NoStop}%
\bibitem [{\citenamefont {Nagaosa}\ \emph {et~al.}(2010)\citenamefont
  {Nagaosa}, \citenamefont {Sinova}, \citenamefont {Onoda}, \citenamefont
  {MacDonald},\ and\ \citenamefont {Ong}}]{Nagaosa2010}%
  \BibitemOpen
  \bibfield  {author} {\bibinfo {author} {\bibfnamefont {N.}~\bibnamefont
  {Nagaosa}}, \bibinfo {author} {\bibfnamefont {J.}~\bibnamefont {Sinova}},
  \bibinfo {author} {\bibfnamefont {S.}~\bibnamefont {Onoda}}, \bibinfo
  {author} {\bibfnamefont {A.~H.}\ \bibnamefont {MacDonald}}, \ and\ \bibinfo
  {author} {\bibfnamefont {N.~P.}\ \bibnamefont {Ong}},\ }\href {\doibase
  10.1103/revmodphys.82.1539} {\bibfield  {journal} {\bibinfo  {journal} {Rev.
  Mod. Phys.}\ }\textbf {\bibinfo {volume} {82}},\ \bibinfo {pages} {1539}
  (\bibinfo {year} {2010})}\BibitemShut {NoStop}%
\end{thebibliography}

%

\clearpage

\pagebreak
\widetext
\begin{center}
\textbf{\large Supplemental Materials: Anomalous Hall effect and magnetic orderings in nanothick V$_5$S$_8$}
\end{center}
\setcounter{equation}{0}
\setcounter{figure}{0}
\setcounter{table}{0}
\setcounter{page}{1}
\makeatletter
\renewcommand{\theequation}{S\arabic{equation}}
\renewcommand{\thefigure}{S\arabic{figure}}
\renewcommand{\bibnumfmt}[1]{[S#1]}
\renewcommand{\citenumfont}[1]{S#1}

\begin{figure}[htbp]
\includegraphics[width=0.9\textwidth]{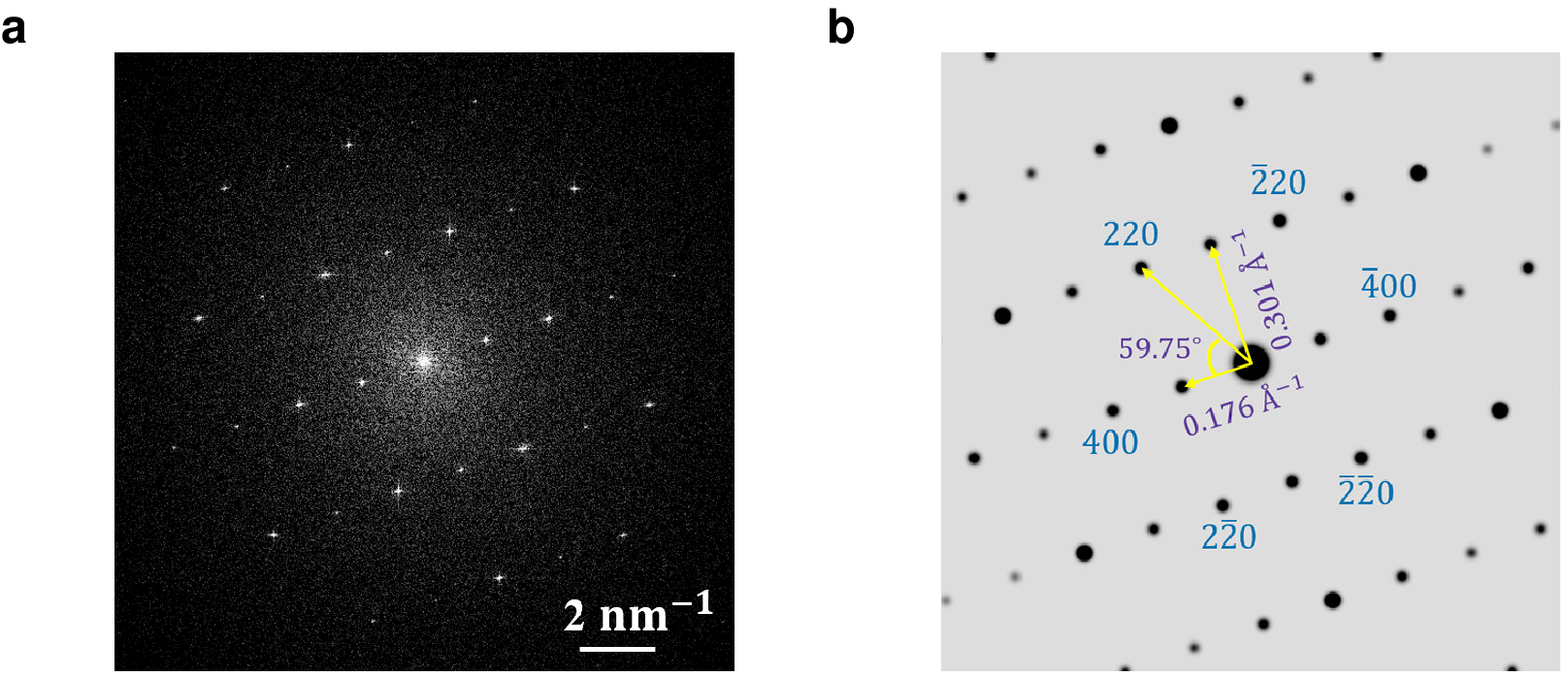}
\caption{Fast Fourier Transformation(FFT) of a high resolution transmission electron microscopic (HRTEM) image. (a) The Reduced FFT of an experimental HRTEM image. (b) The simulated electron diffraction pattern of the ideal V$_5$S$_8$ phase along [001] direction.}
\label{figs1}
\end{figure}
The FFT image of a high resolution TEM image shown in \rfig{figs1} clearly indicates a rectangle arrangement of V atoms. The interplanar spacings of $d_{200}$ and $d_{020}$ are determined to be about $5.75\pm0.05$ $\text \AA$ and $3.35\pm0.05$ $\text \AA$, respectively. \rfig{figs1}b shows the simulated electron diffraction pattern of the ideal V$_5$S$_8$ phase (Fig.~1a). The calculated crystal spacings are about $d_{200}=5.68$ $\text \AA$ and $d_{020}=3.32$ $\text \AA$. The consistent diffractogram between the experimental data and simulation results indicates that the phase of our CVD sample is V$_5$S$_8$. Subsequently, a Fourier mask filtering technique was applied to the TEM image. The processed HRTEM image (Fig.~1b) highlights the atomic structure.

\clearpage

\begin{figure}[htbp]
\includegraphics[width=0.8\textwidth]{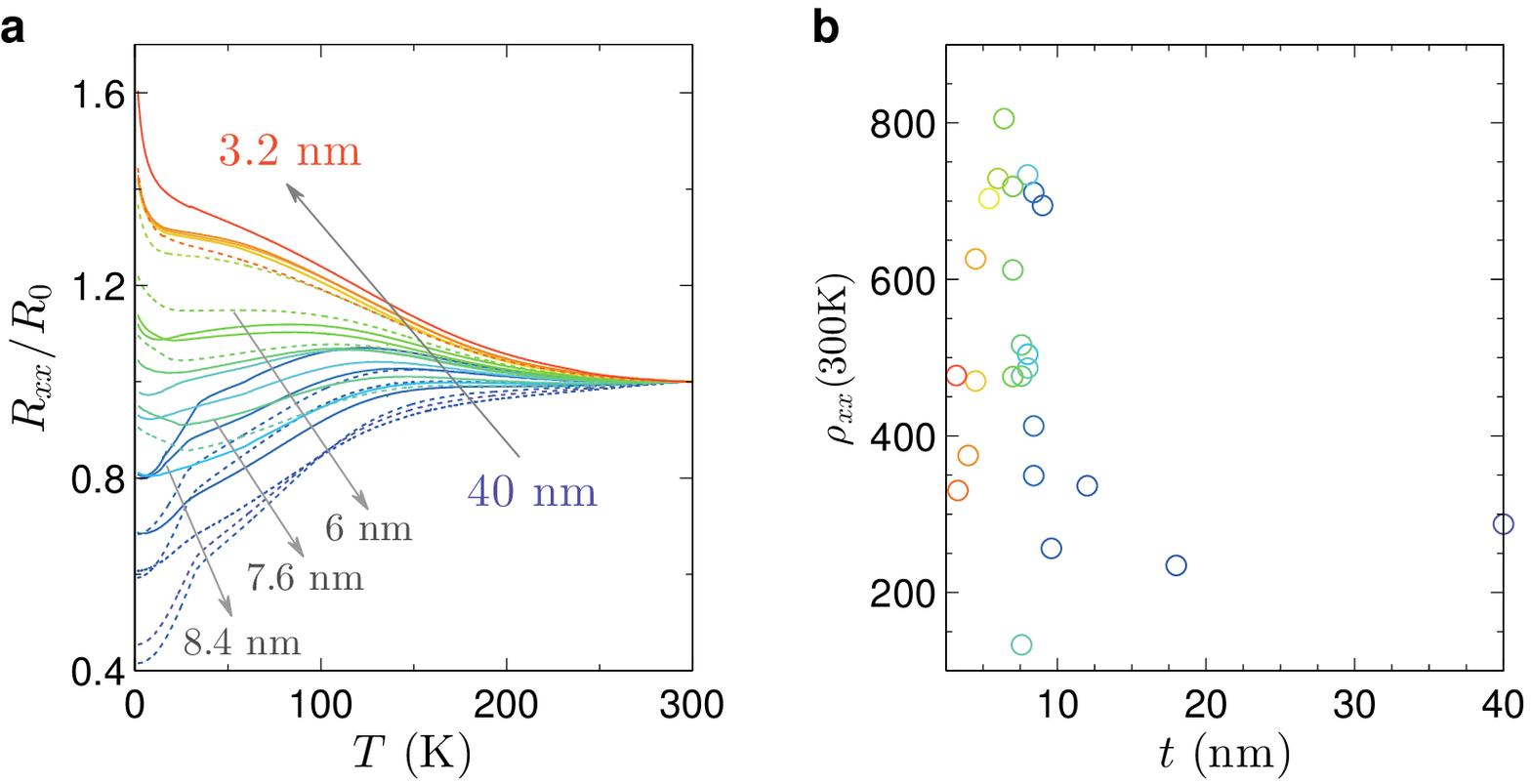}
\caption{(a) Temperature dependence of normalized resistivity. The resistivity is normalized to the 300 K resistivity $R_0$. The thickness of samples ranges from 3.2 to 40 nm. The dashed lines are for as-grown samples, while the solid lines are for peeled-off ones. (b) The corresponding room-temperature resistivity.}
\label{figs2}
\end{figure}
The temperature dependence of resistivity for thick samples is metallic. There is a hump at about 140 K, followed by a sudden drop at 32 K, indication of the antiferromagnetic (AFM) transition. As the thickness is reduced, two trends can be recognized. First, the hump shifts to a lower temperature, so the low temperature resistivity shifts up. Second, the resistivity drop at $T_\text{N}$ turns into an increase (below about 8.4 nm). Although the low-temperature resistivity show enhancement in thinner flakes (below about 6 nm), it remains relatively low down to 3.2 nm. We have measured 40 samples with thickness ranging from 3.2 to 40 nm. The as-grown and further peeled-off samples show similar behavior with reducing thickness. 

\clearpage

\begin{figure}[htbp]
\includegraphics[width=0.6\textwidth]{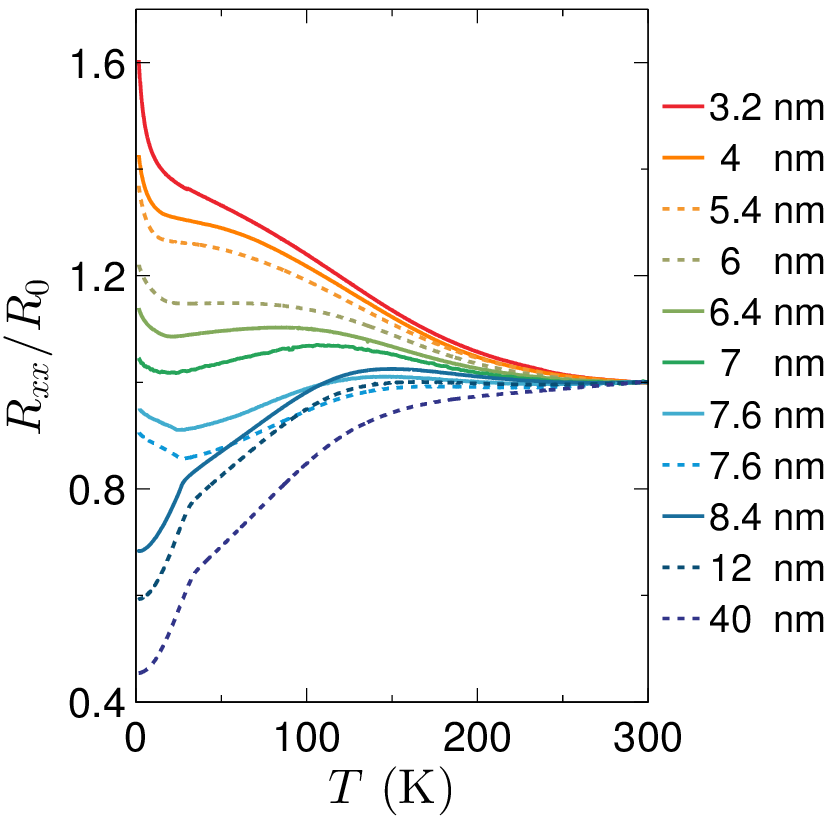}
\caption{Temperature dependence of the normalized resistivity for samples with different thickness. This is a replot of Fig.~1 in the main text. The dashed lines are for as-grown samples, while the solid lines are for peeled-off ones. Both types of samples shows consistent trend upon reducing thickness.}
\end{figure}

\clearpage

\begin{figure}[htbp]
\includegraphics[width=1\textwidth]{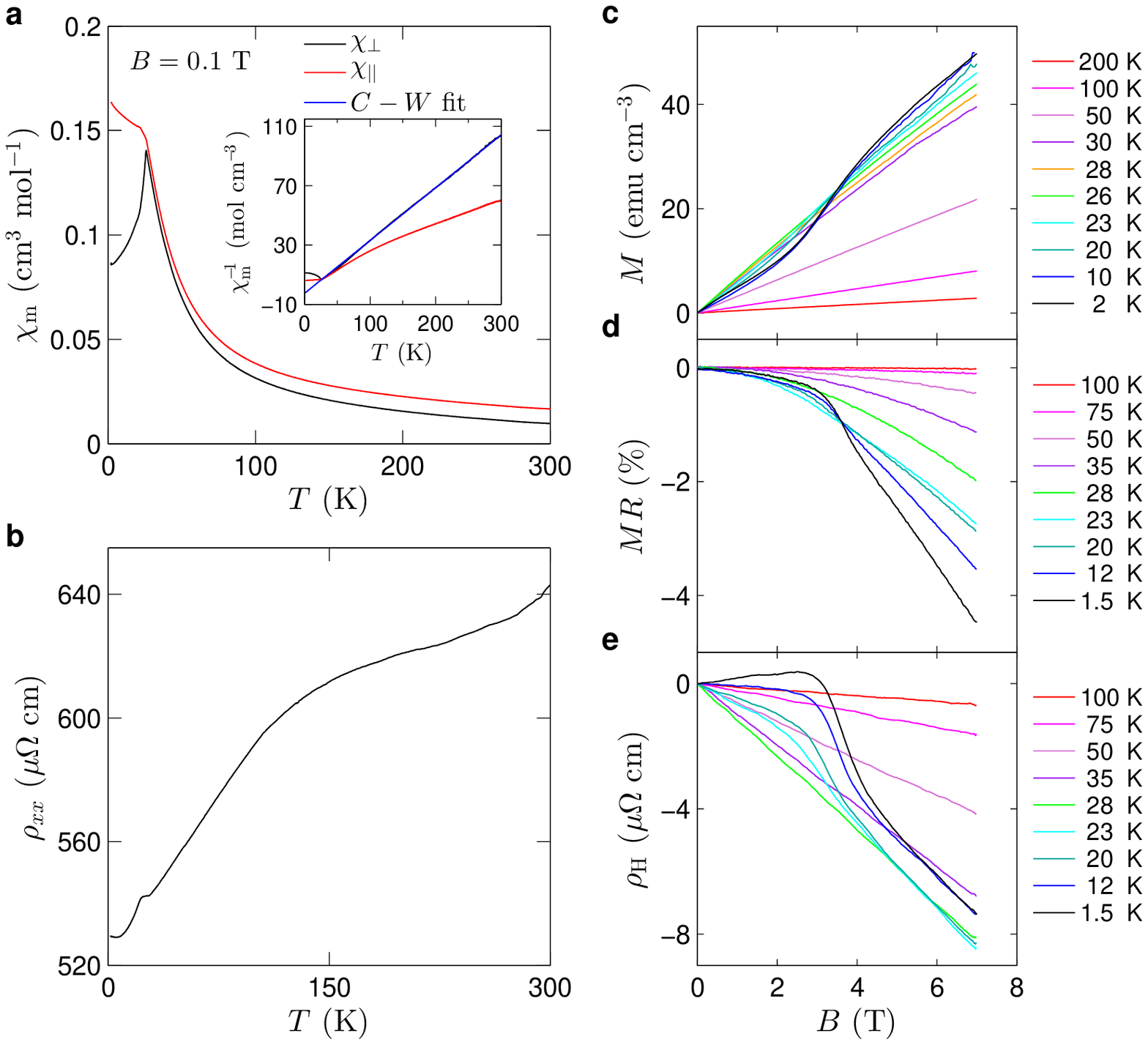}
\caption{Magnetic and transport properties of the bulk V$_5$S$_8$ single crystal. (a) Magnetic susceptibility $\chi_\text{m}$ versus temperature measured at $B=0.1$ T. $\chi_\perp$ and $\chi_\parallel$ represent the susceptibility for out-of-plane and in-plane field, respectively. Inset, Currie-Weiss plot. The blue line is a linear fit of $\chi_\perp^{-1}$. (b) Resistivity versus temperature. Field dependence of (c) magnetization $M$, (d) MR and (e) $\rho_\text{H}$ at different temperatures. $B$ is out-of-plane.}
\label{figs3}
\end{figure}

\clearpage

The magnetic susceptibility $\chi_\text{m}$ is shown in \rfig{figs3}a. Above the AFM transition temperature $T_\text{N}$, $\chi_\text{m}$ follows a Currie-Weiss behaviour. A linear fit of $\chi^{-1}$ yields a Currie-Weiss temperature $\theta_\text{CW}\approx 10$ K. Interestingly, $\theta_\text{CW}> 0$, suggesting a ferromagnetic (FM) interaction, in spite of the AFM transition. Susceptibilities for an out-of-plane and in-plain field, $\chi_\perp$ and $\chi_\parallel$, have been measured. Below $T_\text{N}$, there is a marked difference between $\chi_\perp$ and $\chi_\parallel$. In particular, $\chi_\perp$ is suppressed. The anisotropy of $\chi$ agrees with an AFM state in which spins are aligned out-of-plane. The AFM transition also manifests itself in the $R-T$ curve as a kink. The observations are consistent with earlier studies on V$_5$S$_8$.

In the AFM state, a spin-flop (SF) transition occurs at $B_\text{c}\approx3.6$ T, indicated by the field dependent $M$ in an out-of-plane $B$. An enhancement of susceptibility is observed across the transition. For an in-plane field, $M$ is linear in $B$ in the whole field range. Above $B_\text{c}$, $M$ in out-of-plane and in-plane fields overlap. Correspondingly, the MR exhibits a decrease at $H_\text{c}$. Compared with thick flakes, the MR feature is more smooth and non-zero at low fields, probably due to an averaging effect. $\rho_\text{H}$ of bulk is similar to that of thick flakes, \emph{i.e.} a change of the slope at $H_\text{c}$ and a zero intercept for high field data.

\begin{figure}[htbp]
\includegraphics[width=0.5\textwidth]{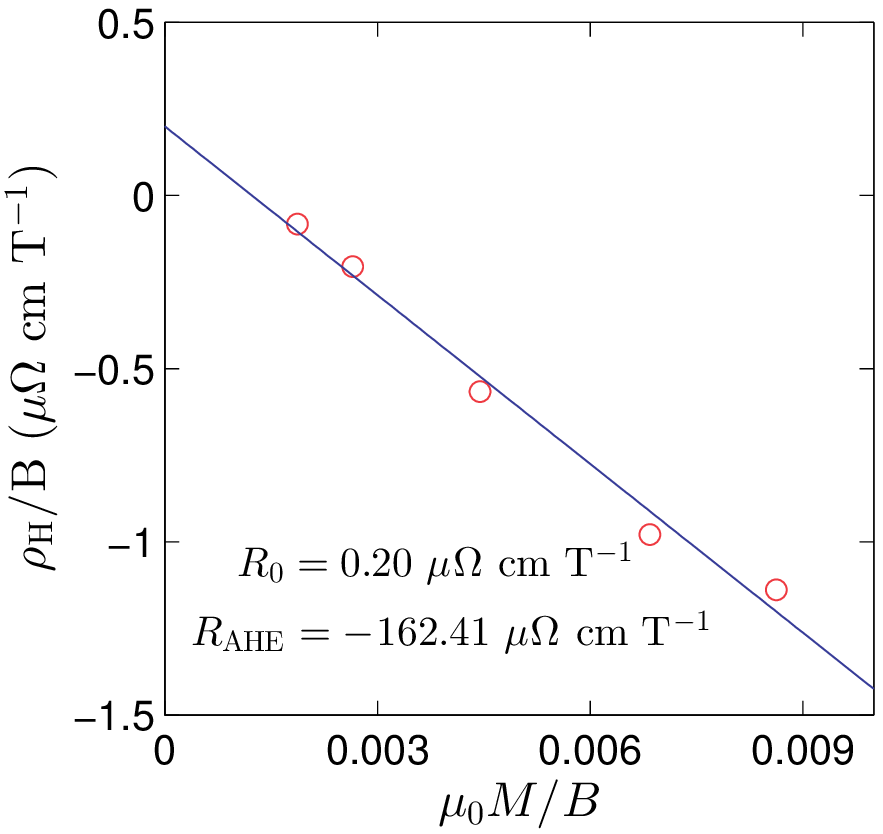}
\caption{Linear relation between $\rho_\text{H}/B$ and $\mu_0 M/B$. The blue line is a linear fit.}
\label{figs4}
\end{figure}

We have postulated that $\rho_\text{H}$ is dominated by the anomalous Hall effect (AHE) from paramagnetic moments, which is proportional to the magnetization, \emph{i.e.} $\rho_\text{AHE}\propto M$. Thus, the total Hall is $\rho_\text{H}(B)=R_0B+R_\text{AHE}\mu_0 M$. In \rfig{figs4}, we plot $\rho_\text{H}/B$ versus $\mu_0 M/B$ at different temperatures above $T_\text{N}$ obtained from \rfig{figs3}c and e and find a good linear relation, which provides strong evidence for our postulation. The linear fit yields $R_0=0.20 $ $\mu\Omega$ cm T$^{-1}$ and $\rho_\text{AHE}=-162.41$ $\mu\Omega$ cm T$^{-1}$. The positive $R_0$ indicates that the carriers are holes, while $\rho_\text{AHE}$ is opposite in sign to $R_0$. The AHE dominates the total Hall above $T_\text{N}$. Below $T_\text{N}$, $\chi_\perp$ declines with decreasing temperature, hence the AHE. So, $\rho_0$ eventually surpasses $\rho_\text{AHE}$, which explains the sign reversal of $\rho_\text{H}$ at the lowest temperature, 1.5 K, in \rfig{figs3}e. For the same reason, $\rho_\text{H}$ becomes negative above the SF transition as $\rho_\text{AHE}$ dominates again.

\clearpage

\begin{figure}[htbp]
\includegraphics[width=0.8\textwidth]{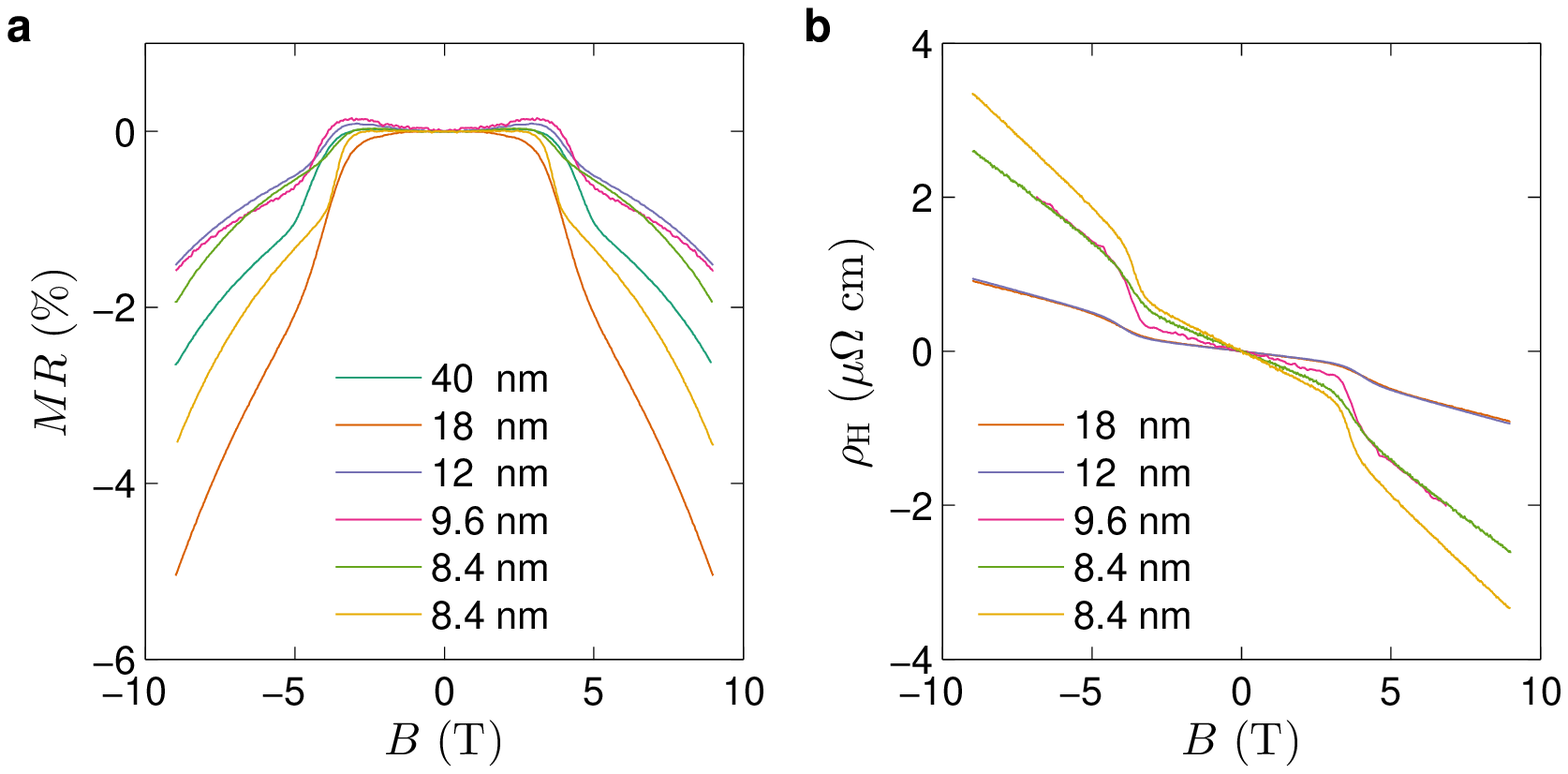}
\caption{Magnetotransport data for thicker flakes in perpendicular field. (a) MR. (b) $\rho_\text{H}$}
\label{figs5}
\end{figure}
Magnetotransport data for more thicker flakes are shown in \rfig{figs5}. The features of the SF transition are prominent. All data are qualitatively similar to the data in the main text. However, the magnitude of the signal displays variations from sample to sample. This is probably due to inhomogeneous intercalation, which gives rise to variations in both local moment density and carrier density.
\clearpage

\begin{figure}[htbp]
\includegraphics[width=0.9\textwidth]{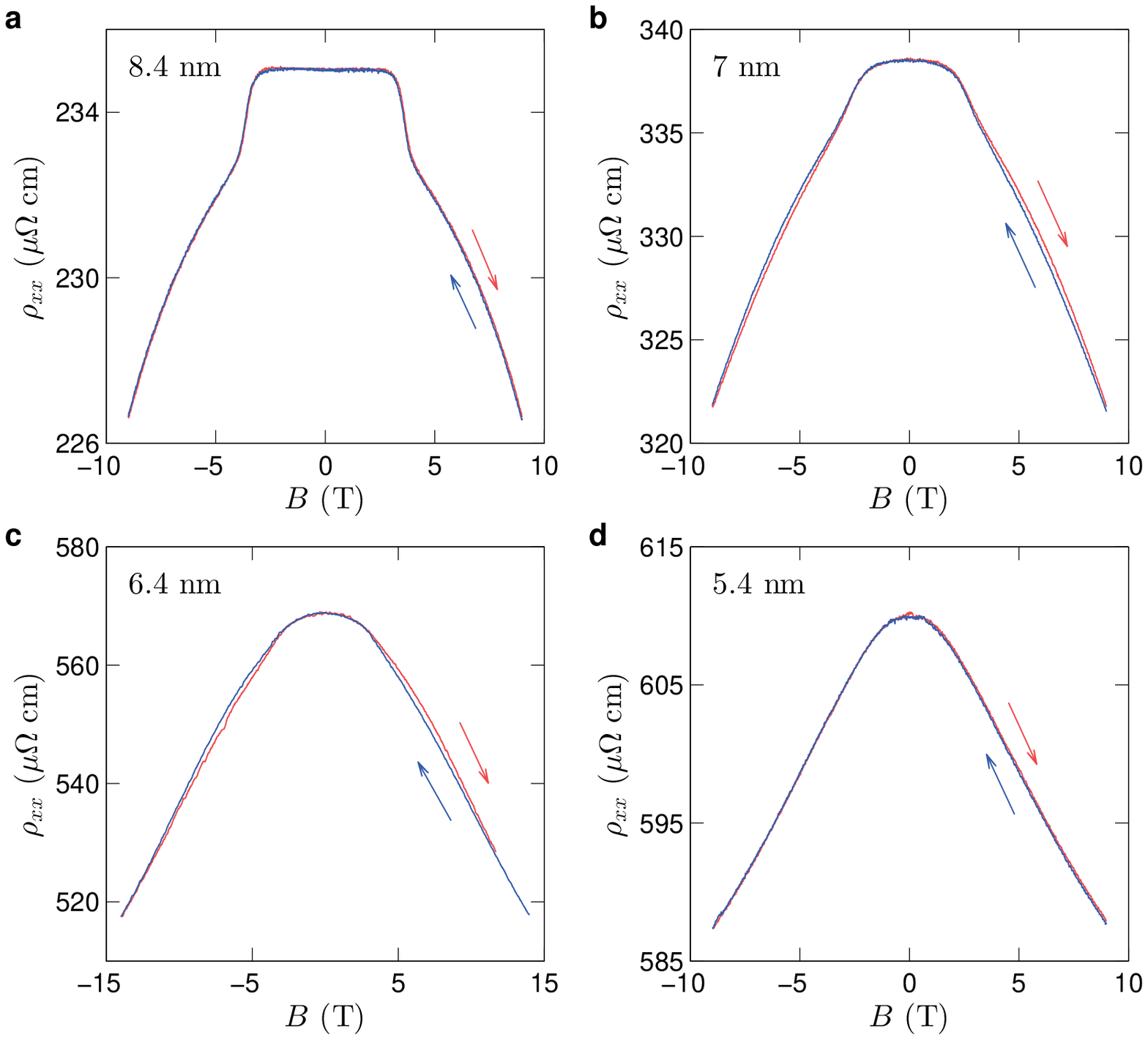}
\caption{MR hysteresis at 1.5 K. MR for flakes of (a) 8.4 nm, (b) 7 nm, (c) 6.4 nm and (d) 5.4 nm. The colored arrows indicate the sweeping direction. No hysteresis can be identified in (a) and (d).}
\label{figs6}
\end{figure}
Flakes of thickness between 8.4 and 5.4 nm often exhibit hysteretic magnetotransport. In \rfig{figs6}, we compare MR of two such samples with other samples. Apparently, there is no appreciable hysteresis in 8.4 and 5.4 nm flakes, while a difference between two field sweeps in 7 and 6.4 nm samples is evident. As we have argued that MR is due to spin-fluctuation scattering, such hysteresis suggests a history dependent spin configuration, for instance, a spin glass state. Formation of such a state here likely results from a competition of magnetic interactions. Magnetization measurements have indeed shown that there are both FM and AFM interactions. The possible spin glass state differs from common ones, as the hysteresis only appears above the SF transition. Therefore, it is a field induced spin glass.

\begin{figure}[htbp]
\includegraphics[width=1\textwidth]{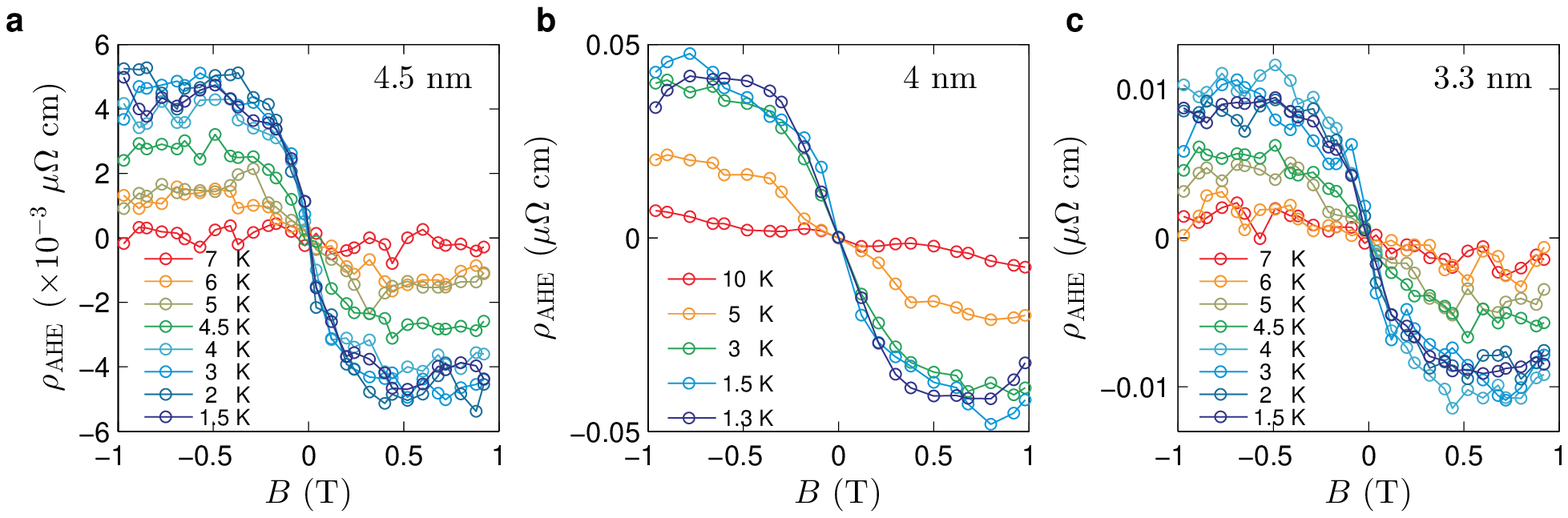}
\caption{Ferromagnetic AHE in thin flakes. (a), (b) and (c) are for three samples with thickness of 4.5, 4 and 3.3 nn, respectively. The linear background has been subtracted.}
\label{figs7}
\end{figure}

\begin{figure}[htbp]
\includegraphics[width=0.6\textwidth]{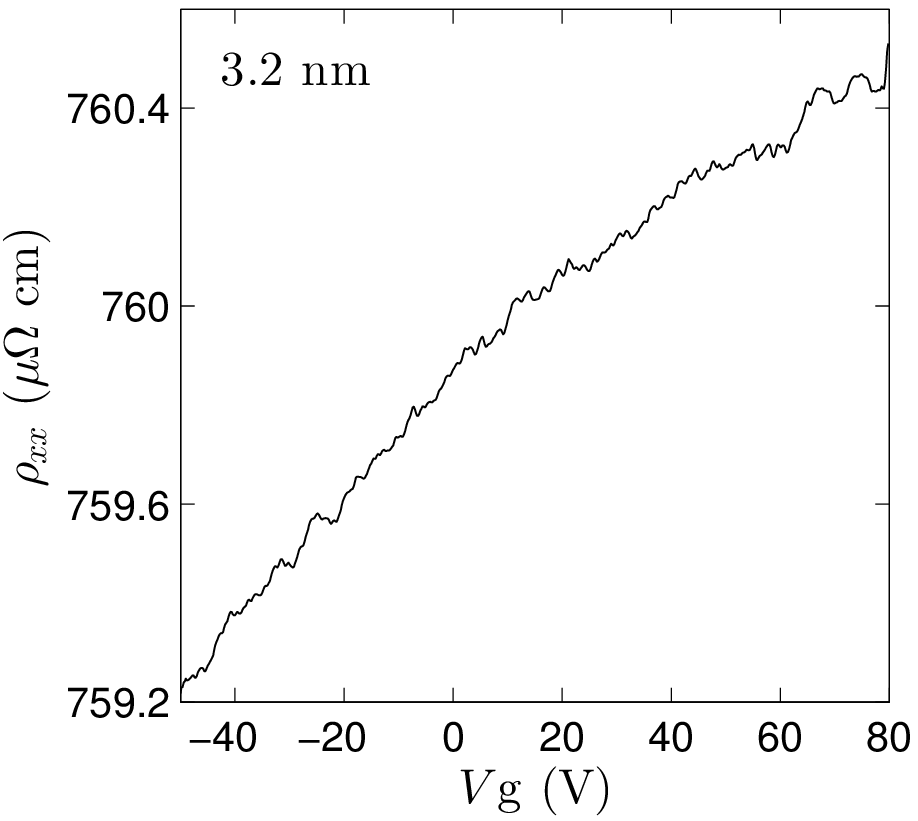}
\caption{Gate dependence of resistivity for a 3.2 nm flake at 1.5 K. The resistivity increases with the gate voltage, suggesting that the carrier type is hole. This is consistent with the analysis of magnetic and transport data for bulk.}
\label{figs8}
\end{figure}

\begin{figure}[htbp]
\includegraphics[width=1\textwidth]{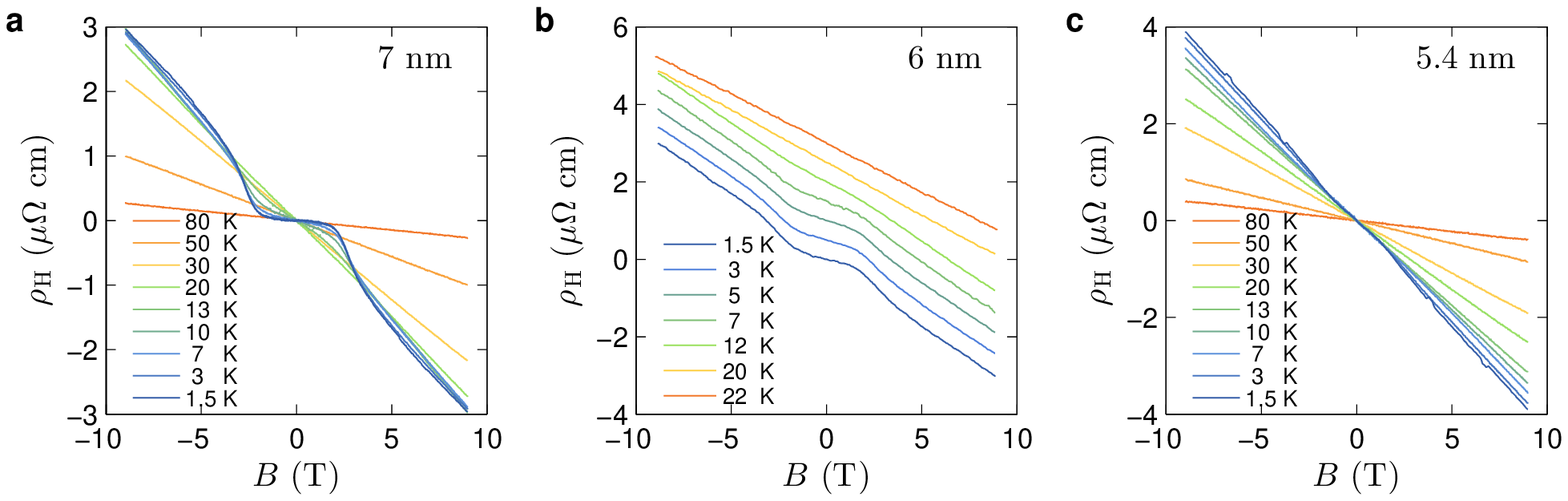}
\caption{Field dependent Hall resistivity for intermediate thickness. (a) 7 nm, (b) 6 nm, and (c) 5.4 nm. Measurements were carried out at a series of temperature, so the transition temperature $T_\text{c}$ at which the nonlinear Hall disappears can be roughly determined.}
\label{figs9}
\end{figure}

\end{document}